\documentclass{jpp}
\usepackage{graphicx}
\usepackage{epstopdf, epsfig}

\usepackage{xcolor}
 \usepackage{amssymb}
 \usepackage{bigints}
 \usepackage{amsmath}
 \usepackage{empheq}
 \usepackage{framed}
 \usepackage{color}
 \usepackage{enumerate}
 \usepackage{ulem}
 \usepackage{bm}
 \usepackage{subcaption}
 \usepackage{appendix}
 \usepackage{multicol}
 \numberwithin{equation}{section}
 \usepackage{upgreek}
 \usepackage{textgreek}
\usepackage{scalerel}
\usepackage{stackengine,wasysym}
\usepackage{stmaryrd} 
\usepackage{xfrac}    
\usepackage{graphicx}
\usepackage{placeins}

\newcommand{\comment}[1]{}

\newcommand{\be}{\begin{equation}}
\newcommand{\ee}{\end{equation}}
\newcommand{\ba}{\[\begin{aligned}}
\newcommand{\ea}{\end{aligned}\]}
\newcommand{\bea}{\begin{eqnarray}}
\newcommand{\eea}{\end{eqnarray}}
\newcommand{\beann}{\begin{eqnarray*}}
\newcommand{\eeann}{\end{eqnarray*}}
\newcommand{\bs}{\begin{split}}
\newcommand{\es}{\end{split}}

\newcommand*{\cE}{\mathcal{E}}
\newcommand*{\cF}{\mathcal{F}}
\newcommand*{\cG}{\mathcal{G}}

\newcommand*{\cJ}{\mathcal{J}}

\newcommand*{\cU}{\mathcal{U}}

\newcommand*{\thv}{\vartheta}

\newcommand*{\ep}{\epsilon}

\newcommand*{\B}{\bm{B}}

\newcommand*{\dl}{\bm{\nabla}}
\newcommand*{\del}{\partial}
\newcommand*{\BD}{\bm{B}\cdot\bm{\nabla}}

\newcommand{\partder}[2]{\frac{\partial #1}{\partial #2}}

\newcommand*{\lbr}{\left(}
\newcommand*{\rbr}{\right)}

\newcommand*{\alphaO}{\alpha^{(0)}}

\newcommand*{\PhiO}{\Phi^{(0)}}

\newcommand*\reallywidetilde[1]{\ThisStyle{%
  \setbox0=\hbox{$\SavedStyle#1$}%
  \stackengine{-.1\LMpt}{$\SavedStyle#1$}{%
    \stretchto{\scaleto{\SavedStyle\mkern.2mu\sim}{.5467\wd0}}{.7\ht0}%
  }{O}{c}{F}{T}{S}%
}}

\def\test#1{$%
  \reallywidetilde{#1}\,
  \scriptstyle\reallywidetilde{#1}\,
  \scriptscriptstyle\reallywidetilde{#1}
$\par}
\makeatletter
\newsavebox{\@brx}
\newcommand{\llangle}[1][]{\savebox{\@brx}{\(\m@th{#1\langle}\)}%
  \mathopen{\copy\@brx\mkern2mu\kern-0.9\wd\@brx\usebox{\@brx}}}
\newcommand{\rrangle}[1][]{\savebox{\@brx}{\(\m@th{#1\rangle}\)}%
  \mathclose{\copy\@brx\mkern2mu\kern-0.9\wd\@brx\usebox{\@brx}}}
\makeatother

\shorttitle{Vacuum magnetic fields with exact quasisymmetry near a flux surface} 

\title{Vacuum magnetic fields with exact quasisymmetry near a flux surface. Part 1: Solutions near an axisymmetric surface}

\shortauthor{W. Sengupta, E. J. Paul, H. Weitzner, A. Bhattacharjee}
\author{Wrick Sengupta\aff{1}\corresp{\email{wrick.sengupta16@nyu.edu}}, Elizabeth J. Paul\aff{2}, Harold Weitzner\aff{1}, Amitava Bhattacharjee\aff{2}}

\affiliation{\aff{1}Courant Institute of Mathematical Sciences, New York University, New York, New York 10012, USA
\aff{2} Department of Astrophysical Sciences, Princeton University, Princeton, NJ,
08543, USA
}

\begin{document}

\maketitle

\begin{abstract}

While several results have pointed to the existence of exactly quasisymmetric fields on a surface \citep{Garren1991a, Garren1991b, Plunk2018}, we have obtained the first such solutions using a vacuum surface expansion formalism. We obtain a single nonlinear parabolic PDE for a function $\eta$ such the field strength satisfies $B = B(\eta)$. Closed-form solutions are obtained in cylindrical, slab, and isodynamic geometries. Numerical solutions of the full nonlinear equations in general axisymmetric toroidal geometry are obtained, resulting in a class of quasi-helical local vacuum equilibria near an axisymmetric surface. The analytic models provide additional insight into general features of the nonlinear solutions, such as localization of the surface perturbations on the inboard side. 
\end{abstract}

\section{Introduction \label{sec:intro}}

Quasisymmetry is a continuous symmetry of the guiding center Lagrangian that, like axisymmetry, implies confinement of collisionless guiding center trajectories near a given flux surface on the drift time scale. Unlike axisymmetry, quasisymmetry does not require a continuous spatial symmetry. Thus three-dimensional quasisymmetric magnetic fields have the potential to possess vacuum rotational transform and improved confinement properties simultaneously. The discovery of quasisymmetry by Boozer \citep{Boozer1983} and the demonstration that quasi-symmetric equilibria could be obtained from asymptotic expansions near the magnetic axis \citep{Garren1991a,Garren1991b} and numerical optimization of MHD equilibria \citep{Nuhrenberg1988} paved the way for the improved confinement of optimized stellarator configurations. More recently, the use of asymptotic expansions near the axis has proved to be fruitful in directly constructing quasisymmetric fields \citep{Landreman2018a,Landreman2018b,Landreman2019,Landreman2019b,Jorge2019} and providing insight into their physical properties \citep{Jorge2020b,Landreman2020b}. 

There remain many open questions related to the existence and nature of quasisymmetric magnetic fields. There is recent evidence that nearly-quasisymmetric fields can be obtained in a volume under the assumption that a small deviation from magnetohydrodynamic (MHD) force balance is introduced \citep{Constantin2020}. However, this result only applies sufficiently close to axisymmetry. By carrying out an asymptotic expansion in the distance near the magnetic axis, it was previously posited that quasisymmetric magnetic fields do not exist in a volume, as an overdetermined system arises at third order for a general axis \citep{Garren1991a, Garren1991b}. If a deviation from MHD force balance is introduced in the form of an anisotropic pressure tensor, then the problem of overdetermination can be avoided, and the series can be continued to high order \citep{Rodriguez2020b, Rodriguez2020c}. However, a solution with volumetric quasisymmetry remains to be obtained. 

While the system of equations for near-axis quasisymmetric magnetic fields with scalar pressure becomes overdetermined, it has been argued that there remains enough freedom in the free functions to achieve exact quasisymmetry on one surface \citep{Garren1991a, Garren1991b, Jorge2020}. A similar plausibility argument was made by demonstrating that an expansion in the distance from axisymmetry can be carried out to all orders to achieve exact vacuum quasi-axisymmetry on a single surface \citep{Plunk2018}. Nevertheless, exact quasisymmetry on a surface which is not asymptotically close to axisymmetry has not been demonstrated previously. 

By expanding in the distance from a flux surface, we obtain vacuum magnetic fields with exact quasisymmetry on one surface. The existence of a surface sufficiently close to quasisymmetry has been shown to improve energetic particle confinement and increase the volume of surfaces with quasisymmetry \citep{Henneberg2019} and may provide an effective transport barrier \citep{Boozer2019}. Obtaining such configurations has previously been quite computationally demanding, requiring numerical optimization of an MHD equilibrium \citep{Spong2001,Drevlak2018} to reduce the harmonics of the field strength in Boozer coordinates that break the desired symmetry \citep{Sanchez2000}. Our approach uses a ``direct construction," like the near-axis or near-axisymmetric expansion techniques, in that numerical optimization of an equilibrium is not required. By directly obtaining solutions with exact quasisymmetry, we additionally gain insight into the nature of the class of solutions.

As we expand in the distance from a single flux surface, our approach uses a formalism similar to local equilibrium models \citep{Hegna2000,Boozer2002,Skovoroda2009,Candy2015}. As noted previously, the local equilibrium equations have a free function of the two angles, which can be used to specify the straight field line angles on a surface \citep{Hegna2000}, the distance to a neighboring magnetic surface \citep{Boozer2002}, or the field strength on a surface \citep{Skovoroda2009}. In this work, we use this freedom to enforce exact quasisymmetry by defining a variable $\eta$ (to be defined below), such that the field strength satisfies $B = B(\eta)$. In this way, Boozer coordinates are not required for the calculation, in contrast with recent work on stellarator vacuum fields near a surface \citep{Boozer2019b}.

Without quasisymmetry, there are three equations for three unknown functions that describe a vacuum field with surfaces, namely the field-line label, scalar potential, and the flux coordinate Jacobian. As quasisymmetry requires an additional constraint, the full global vacuum system with quasisymmetry is overdetermined. However, the near-surface expansion yields enough freedom to enforce quasisymmetry on one surface. As the out-of-surface derivative of the potential does not appear in the lowest order system, the corresponding constraint equation can be replaced by the quasisymmetry constraint in the surface expansion formalism. A limitation of this approach is that a local equilibrium solution is not guaranteed to exist within a global equilibrium. However, as we will discuss in Section \ref{sec:discussion}, the formalism can be extended to look for global quasisymmetric solutions. The near-surface theory presented in this work builds on the formalism developed by Sengupta and Weitzner \citep{Weitzner2016,Sengupta2019,Jaquiery2019}, which was used to study the existence of low-shear magnetic fields with surfaces and construct a global vacuum solution with closed field lines in slab geometry \citep{Weitzner2020}. 



The outline of the paper is as follows. In Section \ref{sec:vacuum_surfaces}, we introduce the equations for vacuum magnetic fields with surfaces, and in Section \ref{sec:quasisymmetry_constraint} we introduce the quasisymmetry constraint. In Section \ref{sec:near_surface}, we expand the relevant equations in the distance from a flux surface, which is used to obtain a single parabolic PDE in Section \ref{sec:parabolic_pde}. Solutions of this equation in slab, cylindrical, and toroidal geometry are presented in Section \ref{sec:QS_solutions}. In Section \ref{sec:numerics}, we demonstrate that numerical solutions of the nonlinear equations can be obtained near a general toroidal axisymmetric surface. These solutions are compared with the analytic models from Section \ref{sec:QS_solutions}, and general features are discussed. Finally, in Section \ref{sec:discussion} we make some concluding remarks.

\section{Vacuum fields with nested surfaces}
\label{sec:vacuum_surfaces}

We consider a generic toroidal coordinate system $(\psi,\theta,\phi)$ where $\psi$ is the toroidal flux and $\theta$ and $\phi$ are the poloidal and toroidal angles that are 2$\pi$ periodic. Denoting the magnetic scalar potential and the field line label by $\Phi$ and $\alpha$ respectively, we represent the vacuum magnetic field with surfaces as
\begin{align}
   \B= \dl \Phi = \dl \psi \times \dl \alpha.
\end{align}
The three components of the above equation yield,
\begin{subequations}
\begin{align}
\sqrt{g}\begin{pmatrix}
\Phi_{,\theta}\\ \Phi_{,\phi}
\end{pmatrix}&=
\begin{pmatrix}
g_{\theta\theta} \quad  \quad g_{\theta\phi}\\ 
g_{\theta\phi} \quad  \quad g_{\phi\phi}
\end{pmatrix}
\begin{pmatrix}
-\alpha_{,\phi} \\ \alpha_{,\theta} 
\end{pmatrix}\\
 \sqrt{g}\Phi_{,\psi} &= g_{\psi\phi} \alpha_{,\theta} - g_{\psi \theta} \alpha_{,\phi}, \label{eq:out_of_surface_vacuum}
\end{align}
\label{eq:vacuum_eqn_phi}
\end{subequations}
where $\sqrt{g} = \left(\dl \psi \times \dl \theta \cdot \dl \phi\right)^{-1}$ is the Jacobian. The first two equations are the ``in-surface" derivatives of $\Phi$, whereas the third equation gives the ``out-of-surface" derivative of $\Phi$. The magnetic differential operator is given by 
\begin{align}
    \BD = \frac{1}{\sqrt{g}}\left( \alpha_{,\theta}\del_\phi - \alpha_{,\phi}\del_\theta\right)=\frac{1}{\sqrt{g}} \{\alpha,\:\:\}_{(\theta,\phi)},
\end{align}
where, $\{f,g\}_{(\theta,\phi)}$ denotes a Poisson bracket of $f$ and $g$ with respect to $(\theta,\phi)$. The field strength is computed from,
\begin{align}
   B^2 = \BD \Phi= \frac{1}{\sqrt{g}}\left(\Phi_{,\phi} \alpha_{,\theta} - \Phi_{,\theta} \alpha_{,\phi}\right).
\end{align}

\section{The quasisymmetry constraint on vacuum fields}
\label{sec:quasisymmetry_constraint}
In the literature various definitions of QS can be found. Here we discuss the following form of QS when the rotational transform is irrational, namely
\begin{align*}
& \text{QS}: \quad  \B\times\dl \psi\cdot \dl B-F(\psi) \BD B =0
\end{align*}
We note that $\BD B$ vanishes whenever $\B \times \dl \psi \cdot \dl B$ does \citep{Landreman2012}, and their ratio remains finite. 
This version of QS has been recently termed ``weak quasisymmetry" \citep{Rodriguez2020a,Burby2020, Constantin2020}. 

In $(\Phi,\psi,\alpha)$ coordinates with $B^2=\dl\Phi\cdot \dl \psi\times\dl\alpha$,  
\begin{align}
    \B\times\dl\psi\cdot \dl=\dl \Phi\times \dl \psi\cdot \dl=B^2 \del_\alpha,\quad\BD = \dl \psi\times \dl \alpha\cdot \dl=B^2 \del_\Phi.
\end{align}
Since $B^2\neq 0$, QS implies 
\begin{align}
    B=B(\Phi +F(\psi)\alpha,\psi).
\end{align}
We define, 
\begin{align}
    \eta= \Phi + F(\psi) \alpha,
\end{align}
such that $B=B(\eta,\psi)$. Note that a similar function can be constructed even if the rotational transform is rational by replacing $\psi$ by $\oint dl/B$ where the integral is carried out along a closed field line \citep{elbarmi_sengupta_weitzner_2020}.
The streamlines of $\eta$ must close on a constant $\psi$ surface. This implies that 
\begin{align}
    \eta = M \theta - N \phi + \widetilde{\eta}(\theta,\phi) = \Phi + F(\psi) \alpha,
\end{align}
where $M$ and $N$ are integers which specify the helicity of the quasisymmetry. We note that $\Phi$ and $\alpha$ also generally have secular terms,
\begin{subequations}
\begin{align}
    \Phi &= \overline{I} \theta + \overline{G} \phi + \widetilde{\Phi}(\theta,\phi) \\
 \alpha &= \theta - \iota \phi + \widetilde{\alpha}(\theta,\phi),
\end{align}
\end{subequations}
where tilde denotes periodic functions. This implies that
$\overline{I} + F(\psi) = M$ and $-\overline{G} + F(\psi)\iota = N$, 
leading to \citep{Helander2014},
\begin{align}
    F(\psi) &= \frac{\overline{G} + N/M \overline{I}}{\iota(\psi) -N/M}.
\label{eq:F_qs}
\end{align}
While $\overline{I}$ vanishes in a global vacuum field, the surface about which we expand may enclose currents which are not in the neighborhood of the surface. As we are performing a local expansion, this does not preclude the existence of currents asymptotically away from the surface. Thus we retain both secular terms in the scalar potential. In terms of the variable $\eta$ one can write $B^2$ as
\begin{align}
    B^2= \frac{1}{\sqrt{g}}\{\alpha,\Phi\}_{\theta,\phi}=\frac{1}{\sqrt{g}}\{\alpha,\eta\}_{\theta,\phi}.
    \label{eq:QS_eqn}
\end{align}
The ``in-surface" derivatives of $\Phi$, $\alpha$, and $\eta$ can then be related to each other through
\begin{align}
\sqrt{g}\begin{pmatrix}
\Phi_{,\theta}\\ \Phi_{,\phi}
\end{pmatrix}&= \sqrt{g}\begin{pmatrix}
\eta_{,\theta} - F \alpha_{,\theta} \\ \eta_{,\phi} - F \alpha_{,\phi}
\end{pmatrix} 
= 
\begin{pmatrix}
g_{\theta\theta} \quad  \quad g_{\theta\phi}\\ 
g_{\theta\phi} \quad  \quad g_{\phi\phi}
\end{pmatrix}
\begin{pmatrix}
-\alpha_{,\phi} \\ \alpha_{,\theta}
\end{pmatrix},
\label{eq:vacuum_eqn}
\end{align}
which yields,
\begin{align}
\sqrt{g}F\begin{pmatrix}
\eta_{,\theta}  \\ \eta_{,\phi}
\end{pmatrix} 
= 
\begin{pmatrix}
\quad g_{\theta\theta} \quad \quad \quad  g_{\theta\phi} + F \sqrt{g}\\ 
\: g_{\theta\phi} - F\sqrt{g} \quad  \quad\quad  g_{\phi\phi}\quad \quad
\end{pmatrix}
\begin{pmatrix}
-F\alpha_{,\phi} \\ +F\alpha_{,\theta}
\end{pmatrix}.
\label{eq:vacuum_eqn2}
\end{align}
This matrix equation can be inverted in order to obtain an equation for $\eta$,
\begin{align}
\frac{\sqrt{g}F}{\mathcal{J}^2 + F^2 \sqrt{g}^2}\begin{pmatrix}
\quad g_{\phi\phi} \quad \quad \quad  -g_{\theta\phi} - F \sqrt{g}\\ 
\: -g_{\theta\phi} + F\sqrt{g} \quad  \quad\quad  g_{\theta\theta}\quad \quad
\end{pmatrix}
\begin{pmatrix}
\eta_{,\theta}  \\ \eta_{,\phi}
\end{pmatrix} 
= \begin{pmatrix}
-F\alpha_{,\phi} \\ +F\alpha_{,\theta}
\end{pmatrix},
\label{eq:vacuum_eqn3}
\end{align}
where we have defined the surface Jacobian to be $\mathcal{J} = \sqrt{g_{\theta\theta} g_{\phi\phi} - g_{\theta\phi}^2}$.

\section{Near-surface expansions}
\label{sec:near_surface}
Let us now expand about a surface $\psi= \psi_0$ such that,
\begin{subequations}
\begin{align}
    \textbf{r}(\psi,\theta,\phi) &= \textbf{r}^{(0)}(\theta,\phi) + \delta \psi\:\textbf{r}^{(1)}(\theta,\phi) + \mathcal{O}(\delta \psi^2) \\
    \alpha(\psi,\theta,\phi) &= \alpha^{(0)}(\theta,\phi) + \delta \psi\:
    \alpha^{(1)}(\theta,\phi) + \mathcal{O}(\delta\psi^2) \\
    \eta(\psi,\theta,\phi) &= \eta^{(0)}(\theta,\phi) + \delta \psi\:
    \eta^{(1)}(\theta,\phi) + \mathcal{O}(\delta\psi^2), 
\end{align}
\end{subequations}
where $\delta \psi = \psi - \psi_0$. Note that we are using a standard Taylor expansion in linear power of $\psi$, as opposed to the expansion in $\sqrt{\psi}$ employed in the near-axis expansions. As we are expanding about a non-degenerate surface away from the axis, the coordinate singularity is avoided and a regular Taylor series suffices.

At $\mathcal{O}((\delta\psi)^0)$ the vacuum equations \eqref{eq:vacuum_eqn_phi} yield,
\begin{subequations}
\begin{align}
  \sqrt{g}^{(1)}
    \begin{pmatrix}
\PhiO_{,\theta}\\ \PhiO_{,\phi}
\end{pmatrix}&=
\begin{pmatrix}
g^{(0)}_{\theta\theta} \quad  \quad g^{(0)}_{\theta\phi}\\ 
g^{(0)}_{\theta\phi} \quad  \quad g^{(0)}_{\phi\phi}
\end{pmatrix}
\begin{pmatrix}
-\alphaO_{,\phi} \\ \alphaO_{,\theta} 
\end{pmatrix}\\
 \sqrt{g}^{(1)} \Phi^{(1)}&= \textbf{r}^{(1)} \cdot \left( \textbf{r}^{(0)}_{,\phi} \alpha^{(0)}_{,\theta} - \textbf{r}^{(0)}_{,\theta} \alpha^{(0)}_{,\phi} \right)
\end{align}
\end{subequations}
where,
\begin{subequations}
\begin{align}
    \sqrt{g}^{(1)}&=  \textbf{r}^{(1)} \cdot \textbf{r}^{(0)}_{,\theta} \times \textbf{r}^{(0)}_{,\phi} \\
    g^{(0)}_{x_ix_j} &= \textbf{r}^{(0)}_{,x_i} \cdot \textbf{r}^{(0)}_{,x_j},
\end{align}
\end{subequations}
for $x_i, x_j \in \{\theta,\phi\}$.

We can define the angles in the neighborhood of the surface such that the metric tensor simplifies \citep{Boozer2002,Imbert2019},
\begin{subequations}
\begin{align}
    \textbf{r}^{(1)} \cdot \textbf{r}_{,\theta}^{(0)} &= \textbf{r}^{(1)} \cdot \textbf{r}_{,\phi}^{(0)} = 0.
\end{align}
\label{eq:local_coordinates}
\end{subequations}
Defining $\rho^{(1)} = |\textbf{r}^{(1)}|$ and $\mathcal{J}^{(0)} = \sqrt{g}^{(1)}/\rho^{(1)}$, we find that with such a choice of coordinates,
\begin{align}
\mathcal{J}^{(0)} = |\textbf{r}^{(0)}_{,\theta} \times \textbf{r}^{(0)}_{,\phi}|.
\end{align}
This eliminates the ``out-of-surface" equation for $\Phi_{,\psi}$ \eqref{eq:out_of_surface_vacuum} locally but not globally. So our vacuum equations simplify to,
\begin{align}
\rho^{(1)}\mathcal{J}^{(0)}\begin{pmatrix}
\PhiO_{,\theta}\\ \PhiO_{,\phi}
\end{pmatrix}&=
\begin{pmatrix}
g^{(0)}_{\theta\theta} \quad  \quad g^{(0)}_{\theta\phi}\\ 
g^{(0)}_{\theta\phi} \quad  \quad g^{(0)}_{\phi\phi}
\end{pmatrix}
\begin{pmatrix}
-\alphaO_{,\phi} \\ \alphaO_{,\theta} 
\end{pmatrix}.
\end{align}
As shown in Appendix \ref{Appendix_geometries}, a global toroidal coordinate system $(\psi,\theta,\phi)$ can also be utilized to bring the ``in-surface" vacuum equations to the same form. In order to carry out a fully global calculation, one must retain the equation for $\Phi_{,\psi}$ \eqref{eq:out_of_surface_vacuum}. In a local expansion, this equation only gives higher-order corrections to $\Phi$ and will not be considered in this work. To enforce QS, we need to rewrite the vacuum equations in terms of the variable $\eta$. Using
\begin{subequations}
\begin{align}
    \Phi^{(0)} &= \eta^{(0)} - F^{(0)} \alpha^{(0)} \\
   \rho^{(1)}\mathcal{J}^{(0)} B^2(\eta^{(0)}) &= \{\alphaO,\PhiO\}_{\theta,\phi}= \{\alphaO,\eta^{(0)} \}_{\theta,\phi},
\end{align}
\end{subequations}
to eliminate $\Phi$, we obtain the following set of equations,
\begin{subequations}
\begin{align}
\begin{pmatrix}
-F^{(0)}\alpha_{,\phi}^{(0)} \\ +F^{(0)}\alpha_{,\theta}^{(0)}
\end{pmatrix} &= \frac{\rho^{(1)}F^{(0)}}{\mathcal{J}^{(0)}\left(1 +  \left(F^{(0)}\rho^{(1)}\right)^2 \right)}\begin{pmatrix}
\quad g_{\phi\phi}^{(0)} \quad \quad \quad  -g_{\theta\phi}^{(0)} - F^{(0)} \rho^{(1)} \mathcal{J}^{(0)}\\ 
\: -g_{\theta\phi}^{(0)} + F^{(0)}\rho^{(1)} \mathcal{J}^{(0)} \quad  \quad\quad  g_{\theta\theta}^{(0)}\quad \quad
\end{pmatrix}
\begin{pmatrix}
\eta_{,\theta}^{(0)}  \\ \eta_{,\phi}^{(0)}
\end{pmatrix} 
\\
    \rho^{(1)}\mathcal{J}^{(0)} B^2(\eta^{(0)}) &= \eta^{(0)}_{,\phi}\alpha^{(0)}_{,\theta}-\eta^{(0)}_{,\theta}\alpha^{(0)}_{,\phi}.
\end{align}
\label{eq:nonlinear_eqns}
\end{subequations}
For the remainder of the text, the superscripts will be dropped for brevity. If the surface is axisymmetric we have,
\begin{align}
    g_{\phi\theta}=0,\quad \cJ=\sqrt{g_{\theta\theta}g_{\phi\phi}}.
\end{align}
Making a coordinate transformation to
\begin{align}
    \Theta = \int d \theta \, \frac{\sqrt{g_{\theta \theta}}}{\sqrt{g_{\phi\phi}}},
\end{align}
we rewrite the vacuum equations and the QS condition as
\begin{subequations}
\begin{align}
\rho
\begin{pmatrix}
\Phi_{,\Theta}\\ \Phi_{,\phi}
\end{pmatrix}&=
\begin{pmatrix}
-\alpha_{,\phi} \\ \alpha_{,\Theta} 
\end{pmatrix}\\
\rho g_{\phi\phi} B^2 &=\{\alpha,\Phi\}_{\Theta,\phi}.
\end{align}
\label{eq:QS_system1}
\end{subequations}
The first two equations have the structure of the generalized Cauchy-Riemann condition and reduces to standard Cauchy-Riemann if $\rho=1$. For the QS problem, $\rho$ must be self-consistently determined along with $\Phi$ and $\alpha$. We note that $\Phi$ and $\alpha$ are therefore pseudo-analytic functions \citep{Bers1953}.

We similarly write the equations in terms of $\eta$ and $\alpha$ \eqref{eq:nonlinear_eqns} with this change of variables as,
\begin{subequations}
\begin{align}
\begin{pmatrix}
-F\alpha_{,\phi} \\ +F\alpha_{,\Theta}
\end{pmatrix} &= \frac{\rho F}{1 +  \left(F\rho\right)^2 } \begin{pmatrix}
    \quad 1 \quad   -\rho F\\ 
    \rho F \quad  \quad 1
    \end{pmatrix}
    \begin{pmatrix}
        \eta_{,\Theta}\\ \eta_{,\phi}
    \end{pmatrix}
\\
    \rho B^2(\eta) G(\Theta) &= \eta_{,\phi}\alpha_{,\Theta}-\eta_{,\Theta}\alpha_{,\phi},
\end{align}
\label{eq:nonlinear_eqns_eta_alpha}
\end{subequations}
where we have defined $G(\Theta) = g_{\phi\phi}$.


 \section{A parabolic PDE to enforce exact QS on a surface}
\label{sec:parabolic_pde}


We now simplify \eqref{eq:nonlinear_eqns_eta_alpha} to get a single PDE for $\eta$. Solving for $\rho F$ in terms of $\eta$ and $B(\eta)$ we obtain,
       \begin{align}
           1+(\rho F)^2= \frac{\eta_{,\Theta}^2+\eta_{,\phi}^2}{G(\Theta)B(\eta)^2}.
       \end{align}
        Eliminating $\alpha$ by equating mixed partial derivatives, we now get a single nonlinear partial differential equation for $\eta$,
       \begin{align}
         (\del^2_{\Theta}+\del^2_{\phi})\eta + \frac{2}{\tan{2\Omega}}(\eta_{,\Theta}\del_\Theta+\eta_{,\phi}\del_\phi)\Omega+2(\eta_{,\Theta}\del_\phi-\eta_{,\phi}\del_\Theta)\Omega =0,
    \end{align}
    where,
    \begin{align}
           \tan\Omega =\rho F= \frac{\sqrt{\eta_{,\Theta}^2+\eta_{,\phi}^2-G(\Theta) B(\eta)^2}}{G(\Theta) B(\eta)^2}.
       \end{align}
    The above equation when further simplified takes the form
       \begin{multline}
           (\rho F \eta_{,\Theta}+\eta_{,\phi})^2\del^2_\phi \eta+ 2(\rho F \eta_{,\Theta}+\eta_{,\phi})(\eta_{,\Theta}-\rho F \eta_{,\phi})\del_\Theta\del_\phi \eta
           + (\eta_{,\Theta}-\rho F \eta_{,\phi})^2\del^2_\Theta \eta
           =\\
           (1+(\rho F)^2)B^3\lbr B'(\eta)(1-(\rho F)^2) +\frac{G'(\Theta)}{2G(\Theta)}\lbr\eta_{,\Theta}(1-(\rho F)^2)-2\rho F \eta_{,\phi}\rbr\rbr.
           \label{eta_eqn}
       \end{multline}
       The right hand side of the $\eta$ equation shows that the ``source" terms are the gradients of the magnetic field strength $B(\eta)$, and the geometry factor $G(\Theta)$. Given these two source terms, $B(\eta)$ and $G(\Theta)$, we can in principle solve the nonlinear $\eta$ equation with the requirement that
       \begin{align}
           \eta = a_1 \Theta -a_2 \phi + \widetilde{\eta}(\Theta,\phi),
           \label{BC}
       \end{align}
    where $\widetilde{\eta}$ is doubly periodic.
    We note that the partial differential equation that $\eta$ satisfies is of parabolic type in general geometry (see Appendix \ref{app:parbolic}) in contrast to the generalized Grad-Shafranov equation \citep{Burby2020} which is elliptic. The repeated characteristic of the parabolic equation is given by
    \begin{align}
        \frac{d\Theta}{d\phi}= \frac{\eta_{,\Theta}-\rho F \eta_{,\phi}}{\rho F \eta_{,\Theta}+\eta_{,\phi}} = -\frac{F\alpha_{,\phi}}{F\alpha_{,\Theta}}.
    \end{align}
    The last expression follows from using the vacuum equations. 
    Therefore, constant $\alpha$ lines are the repeated characteristics for the $\eta$ equation.  
    
    We shall now make a few changes of variables that considerably simplify the nonlinear parabolic equation. Firstly, to bring the parabolic equation to the standard form we shall transform to $(\Theta,F\alpha)$ coordinates. To further simplify the equations, in particular the source terms, we change variables from ($\eta,\Theta$) to ($\cU,\vartheta$), where,
    \begin{align}
        \cU= \int\frac{d\eta}{B(\eta)},\quad \vartheta= \int \: d\Theta\: G(\Theta).
    \end{align}
    The QS equation in these variable reads
    \begin{align}
    \cU_{,\thv\thv}+\cU_{,\thv}\lbr \cU_{,\thv}^2-1\rbr\lbr \frac{G'(\thv)}{2G(\thv)}+\frac{B'(\cU)}{B(\cU)}\frac{\cU_{,\thv}}{1-B\: \cU_{,F\alpha}}\rbr=0.
    \label{QS_eqn}
    \end{align}
    The quantity $\rho$ is given by 
    \begin{align}
        \rho F = \pm\frac{\sqrt{\cU_{,\thv}^2-1}}{1-B\: \cU_{,F\alpha}}.
        \label{rhoF_eqn}
    \end{align}    
The positive and negative signs of $\rho$ ($|\textbf{r}_{,\psi}|$) represents outward and inward surface expansions corresponding to the assumption that the flux increases or decreases away from the axis. Eqns. \eqref{QS_eqn}-\eqref{rhoF_eqn} constitute the principal result of this work. The solution of \eqref{QS_eqn} yields $\eta(\theta,\phi)$ and thus the quasisymmetric field strength through $B(\eta)$. The necessary deformation of the axisymmetric geometry to support quasisymmetric fields is then provided by \eqref{rhoF_eqn}.

We now make a few observations from the structure of \eqref{QS_eqn} and \eqref{rhoF_eqn}. Firstly, for $\rho$ to be real, we must have
\begin{align}
    |\cU_{,\thv}|> 1. 
\end{align}
The case $\cU_{,\thv}=0$ is not allowed since $\rho=0$ implies that the flux surfaces overlap. Physically, $\rho=0$ would correspond to the formation of islands. Secondly, if the gradients of $G$ and $B$ are zero, $\eta=B \cU$ and the general solution of $\cU$ is linear in the angle $\thv$,
\begin{align}
    \cU = u_1(F\alpha)\: \thv + u_2(F\alpha). 
\end{align}
Since $\eta$ must satisfy the boundary condition \eqref{BC}, $u_1(F\alpha)$ must be a constant and $u_2(F\alpha)$ must be a periodic function of $F\alpha$. Furthermore, $u_1\neq 1$ to avoid $\rho=1$. We note that the case $B'=0$ corresponds to local isodynamic fields \citep{Palumbo1968}.

In the next Section, we shall look for exact and approximate analytical solutions of the QS equation in slab, cylindrical, and toroidal geometries. We will show that there are critical differences in the nature of the QS solutions for slab and cylinder geometry for which $G$ is a constant and a real axisymmetric torus for which $G=G(\thv)$.
       
 \section{Solutions of the QS equation in different geometries}  
 \label{sec:QS_solutions}
\subsection{Slab and cylindrical geometries}   
 In slab and cylindrical geometry the geometry factor $G$ is unity (Appendix \ref{Appendix_geometries}). Hence, the only source term in the QS equation is the gradient of $B$. Exact travelling wave solutions to the $\eta$ equation \eqref{eta_eqn} can be constructed such that $\eta=\eta(\Theta-c \phi)$ for constant $c$. The resulting ordinary differential equation (ODE) in the variable $(\Theta-c \phi)$ is,
 \begin{align}
     \eta'' -\frac{B'(\eta)}{B(\eta)}\eta'^2\lbr 2-(1+c^2) \lbr\frac{\eta'}{B}\rbr^2\rbr=0,
 \end{align}
which has an exact solution given by
\begin{align}
\int \frac{d\eta}{k B(\eta)^2}\sqrt{1+(1+c^2)(kB(\eta))^2}=\pm (\Theta-c \phi),
\label{eq:traveling_wave}
\end{align} 
where $k$ is the integration constant. Since $B(\eta)$ is a periodic function of $\eta$, the integral on the left defines a function which has periodic terms in $\eta$ and at most a linear secular term in $\eta$ arising from the average of the integrand. This can be readily seen by Fourier expanding the integrand in $\eta$. Therefore, when the function of $\eta$ is inverted, we get a solution which satisfies the boundary condition \eqref{BC}.

Similarly, there exists exact solutions of the form $\cU=\cU(\thv-c F\alpha)$ to the QS equation \eqref{QS_eqn} which reduces to the following ODE in the variable $(\thv-c F\alpha)$,
\begin{align}
    \cU''+\cU'\lbr {\cU'}^2-1\rbr\lbr \frac{B'(\cU)}{B(\cU)}\frac{\cU'}{1+c B\: \cU'}\rbr=0.
    \label{G1_TW_Falplha}
    \end{align}
 The ODE has the exact solution
 \begin{align}
     -\int d\cU \frac{ \pm\sqrt{k(1-(c  B(\cU))^2)+1}+(c B(\cU))^2}{c B(\cU) \left(\pm\sqrt{k \left(1-(c B(\cU))^2\right)+1}+1\right)}= \thv-c F\alpha,
 \end{align}
 with $k$ as an integration constant. We observe that there are two possible travelling wave solutions in each of the above cases due to quadratic nature of the equations. This can be interpreted as solutions for which the toroidal flux increases in the direction toward or away from the axis. The existence of two linearly independent solutions is also obtained in the system of equations that arises in an asymptotic expansion near axisymmetry \citep{Plunk2018,Plunk2020}.
 
 \subsection{Toroidal geometry with arbitrary aspect ratio and arbitrary shaping}
 \label{sec:toroidal_solutions}
 
 If $G'(\thv)\neq 0$, exact travelling wave solutions do not exist since $G(\thv)$ breaks the travelling wave symmetry. We shall treat the QS equations in this case numerically in Section \ref{sec:numerics}. However, analytical progress can still be made in several interesting cases. In the following we shall give the analytical details of these cases to provide physical insight into the nature of the QS solutions and also to serve as benchmarks for the numerical solution.  
 
\subsubsection{Complete axisymmetry}\label{complete_A}
The complete axisymmetric case for arbitrary aspect ratio and shaping can be tackled analytically. In this limit we have $ G=G(\thv)$, $\cU= \cU( \thv)$, $B=B( \cU)$, and $\rho=\rho(\thv)$. The QS equation reduces to,
         \begin{align}
              \cU''+ \cU'( \cU'^2-1)\lbr \frac{G'}{2G}+\frac{B'}{B} \cU' \rbr=0.
         \end{align}
The above equation can be integrated once and we obtain
\begin{align}
     G B^2  \lbr 1- \frac{1}{ \cU'^2} \rbr = c_0,
     \label{eq:axisymmetric_constant}
\end{align}
where $c_0$ is the integration constant. We can rewrite the above equation in the form of an energy equation with a potential $V(\cU,\thv)$,
        \begin{align}
        \frac{1}{2} \cU'^2+V( \cU, \thv)=0, \quad V( \cU, \thv)= \frac{1}{2}\frac{G B^2}{c_0-G B^2}.
        \end{align}  
From \eqref{rhoF_eqn} it follows that
        \begin{align}
            \rho F = \pm \frac{\sqrt{c_0}}{\sqrt{G B^2-c_0}}.
            \label{eq:rhoF_axisymmetric}
        \end{align}
\subsubsection{Isodynamic fields ($B=B_0,B'=0$)}
\label{Isodynamic_B}
The QS equation reduces to the following ODE in the variable $\thv$ when $B=B_0$ is constant on the flux surface,
\begin{align}
    \cU''+\cU'({\cU'}^2-1)\frac{G'}{2G}=0,
\end{align}
which has an exact solution
\begin{align}
    \cU = \pm \int d\thv \:\sqrt{\frac{G(\thv) B^2_0}{G(\thv) B^2_0-c_0}} + \Gamma F\alpha+ \overline{\cU}(F\alpha).
    \label{exact_isodynamic}
\end{align}
The function $\overline{\cU}(F\alpha)$ is an arbitrary periodic function of $F\alpha$ and $\Gamma$ is a constant. The corresponding value of $\rho F$ is 
\begin{align}
    \rho F = \pm \sqrt{\frac{c_0}{G B^2_0-c_0}}\lbr 1-B_0(\Gamma+\overline{\cU}'(F\alpha)) \rbr^{-1}
\end{align}

Palumbo's original isodynamic solution \citep{Palumbo1968} was axisymmetric, but locally we can always find non-axisymmetric isodynamic fields provided $\Gamma\neq 0$ and $\overline{\cU}(F\alpha)$ is not a constant. 
 
 \subsubsection{Localization near $B'=0$ }
 \label{Near_Bp0}
We shall now look at the behavior of a nontrivial non-axisymmetric solution in the neighborhood of $B'=0$ which we assume happens along $\cU=\cU_m$. We shall restrict ourselves to the region $|\delta \cU|=|\cU-\cU_m|\ll 1$. Even though $\delta U\ll 1$, we do not expect $\delta U'$ to be small. Therefore, we can have sharp variations in the solution near $B'=0$.

In the region near $B'=0$, we expect the solution to be close to the isodynamic case given by \eqref{exact_isodynamic}. We therefore make the ansatz that $\cU$ is of the form
\begin{align}
    \cU=\cU_m + \Gamma F\alpha + h(\thv),
\end{align}
where $\Gamma$ is a constant.
Taylor expanding $B(\cU)$ in powers of $\delta \cU$ and keeping only up to the quadratic power in $\delta \cU$ such that $ B=B_0 + B_2  (\delta \cU)^2+\dots$, and keeping all nonlinearities in $\delta\cU'$, we get 
    \begin{align}
        \delta\cU''+ \delta\cU'( (\delta\cU')^2-1)\lbr \frac{G'}{2G}+2\frac{B_2 \delta\cU}{B_0}\frac{ \delta\cU'}{1-B_0\Gamma} \rbr=0,
    \end{align}
    where prime indicates differentiation with respect to $\vartheta$. The above equation can be integrated once such that
        \begin{align}
            G B_0^2\lbr 1- \frac{1}{ \delta \cU'^2} \rbr \exp{\lbr\frac{2B_2}{B_0}\frac{ \delta \cU^2}{1-B_0 \Gamma}\rbr}  = c_0,
        \label{eq:c0_near_isodynamic}
        \end{align} 
        We can cast the above expression in a energy equation as follows,
        \begin{align}
            \frac{1}{2} \delta \cU'^2+V( \delta\cU, \thv)=0, \quad V( \delta\cU, \thv)= \frac{1}{2}\lbr\frac{c_0}{B_0^2 G}e^{-\lbr\frac{2B_2}{B_0}\frac{ \delta \cU^2}{1-B_0 \Gamma}\rbr}-1\rbr^{-1}.
        \end{align}      
        We shall show later that the solution has sharp peaks near the minimum of the magnetic field strength, and we compare with the numerically obtained solution in Section \ref{sec:numerics}.

 \subsubsection{Near-isodynamic limit}
 We shall now briefly discuss a bifurcation that can happen when the background geometry $G(\thv)$ and magnetic field strength $B_0(\thv)$ are axisymmetric. If we now look for a small deformation such that magnetic field is QS i.e. $B=B(\eta)$, we can have the following situations:
 \begin{enumerate}
    \item   Both the field strength and $\rho$ are axisymmetric  \label{A}
     \item  Both the field strength and $\rho$ are non-symmetric. \label{QH}
 \end{enumerate}

Case \eqref{A} implies that $\eta=\eta(\thv)$ and $\rho =\rho(\thv) $. We have analyzed this case completely in \ref{complete_A}. Case \eqref{QH} implies that $\eta$ must have dependence on both $\thv$ and $F\alpha$, so that $\rho F$, given by \eqref{rhoF_eqn}, is non-symmetric. We now look for solutions of the QS equation \eqref{QS_eqn} in the form 
\begin{subequations}
 \begin{align}
     \cU&=\cU_0(\thv,F\alpha)+\ep\cU_1(\thv,F\alpha)+\mathcal{O}(\epsilon^2) \\
     B&=B( \cU)=B_0+\ep B_1(\cU)+\mathcal{O}(\epsilon^2)\\
     \rho&=\rho_0+\ep \rho_1+\mathcal{O}(\epsilon^2),
 \end{align}
 \end{subequations}
 where $B_0$ is a constant background field. The expansion parameter $\ep$ is similar to the large aspect ratio parameter, however we do not expand the geometry factor. We further require that to lowest order, $\cU$ depends on $F\alpha$ only through a secular term $\Gamma F\alpha$,
 \begin{align}
    \cU_0(\thv,F\alpha)= \Gamma F\alpha +h_0( \thv).
    \label{U0}
 \end{align}
 For QA we must have $\Gamma=0$. 
 
 The lowest order calculations are analogous to the case described in Section \ref{Near_Bp0}, since $B=B_0$ is constant to this order. Denoting $\thv$ derivatives by primes, \eqref{QS_eqn} to lowest order implies
\begin{align*}
h''_0+h'_0({h'_0}^2-1) \frac{G'}{2G}=0,
\end{align*}
which leads to the energy equation
\begin{align*}
          \frac{1}{2}{h'_0}^2+V(h_0, \thv)=0, \quad V(h_0, \thv)= \frac{1}{2}\frac{G B_0^2}{c_0-G B_0^2},
\end{align*}      
        where, the integration constant $c_0$ is a periodic function of $F\alpha$. It is convenient to separate $c_0$ into a constant piece $C_0$ and a periodic part $\widetilde{c}_0$.
The quantity $\rho_0 F$ is given by
\begin{align}
    \rho_0 F = \sqrt{\frac{c_0}{G B_0^2-c_0}}\frac{1}{1-B_0\Gamma}.
    \label{eq:rho_isodynamic}
\end{align}
 
To $O(\ep)$, we obtain the following linear equation for $\cU_1$ 
\begin{align}
\lbr \cU'_1\sqrt{\frac{(G B^2_0-c_0)^3}{G B^2_0}}\rbr'=\frac{c_0}{1-B_0\Gamma}\frac{B'_1(\cU_0)}{B_0}\sqrt{\frac{GB_0^2}{GB^2_0-c_0}}.
    \label{U1_eqn}
\end{align}
We note that since $\cU'_1$ is periodic and $(B_0,\Gamma,c_0)$ are all independent of $\thv$, the right hand side of \eqref{U1_eqn} must satisfy the condition,
\begin{align}
    \oint d\thv \frac{B'_1(\cU_0)}{B_0}\sqrt{\frac{GB_0^2}{GB^2_0-c_0}}=0.
\end{align}
The above constraint is satisfied if both $B_1$ and $G$ are stellarator symmetric, for example. 
Solving for $\cU_1$ we get
\begin{align}
    \cU_1=c_1(F\alpha)\sqrt{\frac{G B^2_0}{(G B^2_0-c_0)^3}}+\frac{c_0}{1-B_0\Gamma}\int d\thv \frac{B'_1(\cU_0)}{B_0}\sqrt{\frac{GB_0^2}{GB^2_0-c_0}},
\end{align}
where, $c_1$ is a function of $F\alpha$ to be determined by going to the next order. 

The second order equation is once again of the form \eqref{U1_eqn} and the periodicity requirement on $\cU_2$ leads to a Riccati equation for $c_1$. The calculation is straightforward and we shall skip the details. The Riccati equation for $c_1$ is of the form
\begin{align}
    \frac{d c_1}{d(F\alpha)}+2 a_1 c_1 + a_2 c_1^2+a_3=0,
    \label{Riccati_for_c1}
\end{align}
where $a_1$, $a_2$, and $a_3$ are complicated functions of $c_0(F\alpha)$. The function $c_1$ has to be a periodic function of $F\alpha$. Therefore, we must have
\begin{align}
\oint dF\alpha \lbr 2 a_1 c_1 + a_2 c_1^2+a_3 \rbr=0.
\label{avg_Riccati}
\end{align}
Assuming that $c_0= C_0 + \widetilde{c}_0$, we can determine $C_0$ using \eqref{avg_Riccati} for a given periodic function $\widetilde{c}_0$. We note that if $\widetilde{c}_0 =0$ , i.e. $c_0$ is a constant, the Riccati equation \eqref{Riccati_for_c1} has constant coefficients and can be solved exactly as
\begin{align}
c_1=\frac{a_1-\Delta_0 \tan \left(\Delta_0(F\alpha-F\alpha_0)\right)}{a_2}, \quad \Delta_0=\sqrt{a_2 a_3-a_1^2}.
    \label{tan_solution}
\end{align}
Thus, we see that unless $\Delta_0=0$, $c_1$ is not periodic when $c_0$ is a constant. When $\Delta_0=0$, we obtain an axisymmetric solution. 

\subsubsection{A model for QS fields near an axisymmetric surface}
The previous analysis suggest a strong correlation between $\rho F$ and the product of the geometry factor, $G$, and the magnetic field strength, $B^2$. We construct a simple model of QS that captures most of the interesting properties for a large class of travelling-wave like solutions for arbitrary geometry and field strength factors,
\begin{subequations}
 \begin{align}
    \cU &\approx \pm \int d\thv \:\sqrt{\frac{G B^2}{G B^2-c_0}} + \Gamma F\alpha \\
    \rho F &\approx \pm \sqrt{\frac{c_0}{G B^2-c_0}}\lbr 1-B \Gamma \rbr^{-1}.
\end{align}
\label{model_TW_solution}
\end{subequations}
This model is motivated by the solutions obtained in the near-isodynamic limit \eqref{U0}-\eqref{eq:rho_isodynamic}.
In Section \ref{sec:numerics}, we verify this model numerically.
 
\section{Numerical solutions of the QS equations}
\label{sec:numerics}

Although full toroidal geometry can be treated analytically in a few limiting cases, we obtain numerical solutions for the general case. To do so, we construct a spectral solution for $\alpha$, $\eta$, and $\rho$,
\begin{subequations}
\begin{align}
    \alpha &= \theta - \iota \phi + \sum_{i=1}^{N_{\text{modes}}} \alpha_{m_i,n_i} \sin(m_i \theta- n_i \phi)\\
    \eta &= \theta - N/M \phi + \sum_{i=1}^{N_{\text{modes}}} \eta_{m_i,n_i} \sin(m_i \theta - n_i \phi) \\
    \rho &= \sum_{i=1}^{N_{\text{modes}}+1}\rho_{m_i,n_i}\cos(m_i\theta - n_i\phi),
\end{align}
\label{eq:spectral}
\end{subequations}
where we have assumed stellarator symmetry and chosen a normalization such that the secular term in $\eta$ is unity. We note that $N/M$ must be a rational number, which determines the helicity of the quasisymmetry. A nonlinear set of equations is obtained by inserting \eqref{eq:spectral} into \eqref{eq:nonlinear_eqns} and integrating against a set of $N_{\text{modes}}+1$ basis functions,
\begin{align}
    \{x_{m_j,n_j}\} = \{\cos(m_j \theta - n_j \phi)\}.
\end{align}
The set of $3N_{\text{modes}}+3$ unknowns is taken to be $\{\iota,\alpha_{m,n},F,\eta_{m,n},\rho_{m,n}\}$ while the helicity $N/M$ is prescribed in addition to the functional form of $B(\eta)$. The resulting nonlinear set of equations is solved with a trust region method with a prescribed analytic Jacobian matrix. In the following Sections, we verify the numerical solutions in several limits. 

Once the solution is obtained, the Boozer angles can be computed as,  
\begin{subequations}
\begin{align}
    \vartheta_B &= \theta + \frac{-N/M \widetilde{\alpha} + \iota \widetilde{\eta} }{\iota - N/M} \\
    \varphi_B &= \phi + \frac{\widetilde{\eta}  - \widetilde{\alpha}}{\iota - N/M},
\end{align}
\end{subequations}
such that only secular terms remain in the potentials,
\begin{subequations}
\begin{align}
    \Phi(\vartheta_B,\varphi_B) &= \overline{G} \varphi_B + \overline{I} \vartheta_B \\
    \alpha(\vartheta_B,\varphi_B) &= \vartheta_B - \iota \varphi_B.
\end{align}
\end{subequations}

\subsection{Complete axisymmetry}

\begin{figure}
    \centering
    \includegraphics[width=0.8\textwidth]{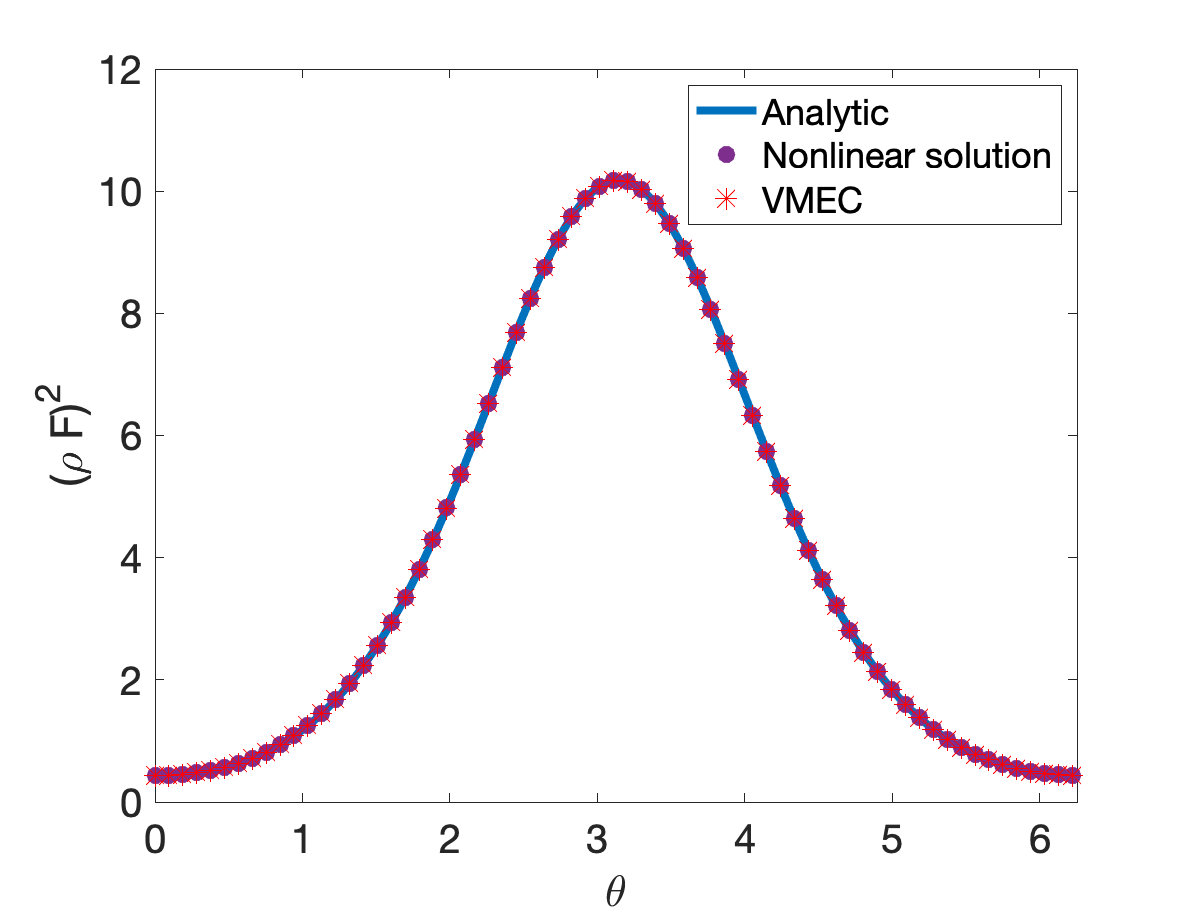}
    \caption{Benchmark of pure axisymmetric solution \eqref{eq:rhoF_axisymmetric}.}
    \label{fig:axisymmetric_x1}
\end{figure}

In the case of pure axisymmetry, we can confirm that solutions of \eqref{eq:nonlinear_eqns} agree with numerical solutions of the Grad-Shafranov equation on a surface. To do so, we compute an axisymmetric equilibrium with the VMEC code \citep{Hirshman1983}. We consider a boundary given by,
\begin{subequations}
\begin{align}
    R(\theta,\phi) &= 3 + \cos(\theta) \\
    Z(\theta,\phi) &= \sin(\theta),
\end{align}
\label{eq:boundary_tok}
\end{subequations}
and impose toroidal current and pressure profiles which are peaked near the axis,
\begin{subequations}
\begin{align}
I_T'(\psi/\psi_0) &= I_0 e^{-(\psi/\psi_0)^2/0.05^2} \\
p(\psi/\psi_0) &= p_0 e^{-(\psi/\psi_0)^2/0.05^2},
\end{align}
\end{subequations}
to provide a non-zero rotational transform. The magnetic field on the surface at $\psi/\psi_0 = 0.995$ is used to construct $\rho$, $\Phi$, $\alpha$, and $\eta$ using \eqref{eq:F_qs} with $N = 0$ and $M = 1$. A Fourier transform is then performed to obtain the functional form of $B(\eta)$. This solution is provided as an initial guess to the numerical solver of the discretized equations \eqref{eq:nonlinear_eqns}, including modes with $m \le 20$, which converges in 3 iterations to a function tolerance of $10^{-8}$. Given the numerical solution, we compute the integration constant $c_0$ from \eqref{eq:axisymmetric_constant}. The relative variation of $c_0$ with respect to $\theta$ is $3.5\times10^{-9}$, and its average over $\theta$ is used to evaluate \eqref{eq:rhoF_axisymmetric}. The value of $\rho F$ computed in this way is displayed as the ``analytic" result in Figure \ref{fig:axisymmetric_x1}. This is compared with the value computed from the VMEC equilibrium and that obtained from the nonlinear solution. As can be seen, the three solutions are in good agreement. The averaged relative error,
\begin{align}
     \text{error} = \frac{\int d \theta \, \left|  (\rho F)^2 - \left((\rho F)^{\text{model}}\right)^2 \right| }{\int d \theta  \, (\rho F)^2},
\end{align}
between the VMEC and nonlinear solution is $2.49\times 10^{-5}$, and the averaged residual between the analytic and nonlinear solution is $1.27\times 10^{-8}$.

\subsection{Bifurcated solutions}

We now consider solutions near an axisymmetric solution for which the field strength and $\rho$ are non-axisymmetric. In order to obtain such numerical solutions, we provide an initial guess, which is a traveling wave solution, $\eta(c\Theta -N/M \phi)$. Here the constants $F$ and $c$ defining the initial condition are chosen such that the secular terms in $\theta$ are unity as in \eqref{eq:spectral} and $N/M$ is taken to match the desired helicity.

\begin{figure}
    \centering
    \begin{subfigure}[b]{0.49\textwidth}
    \includegraphics[width=1.0\textwidth]{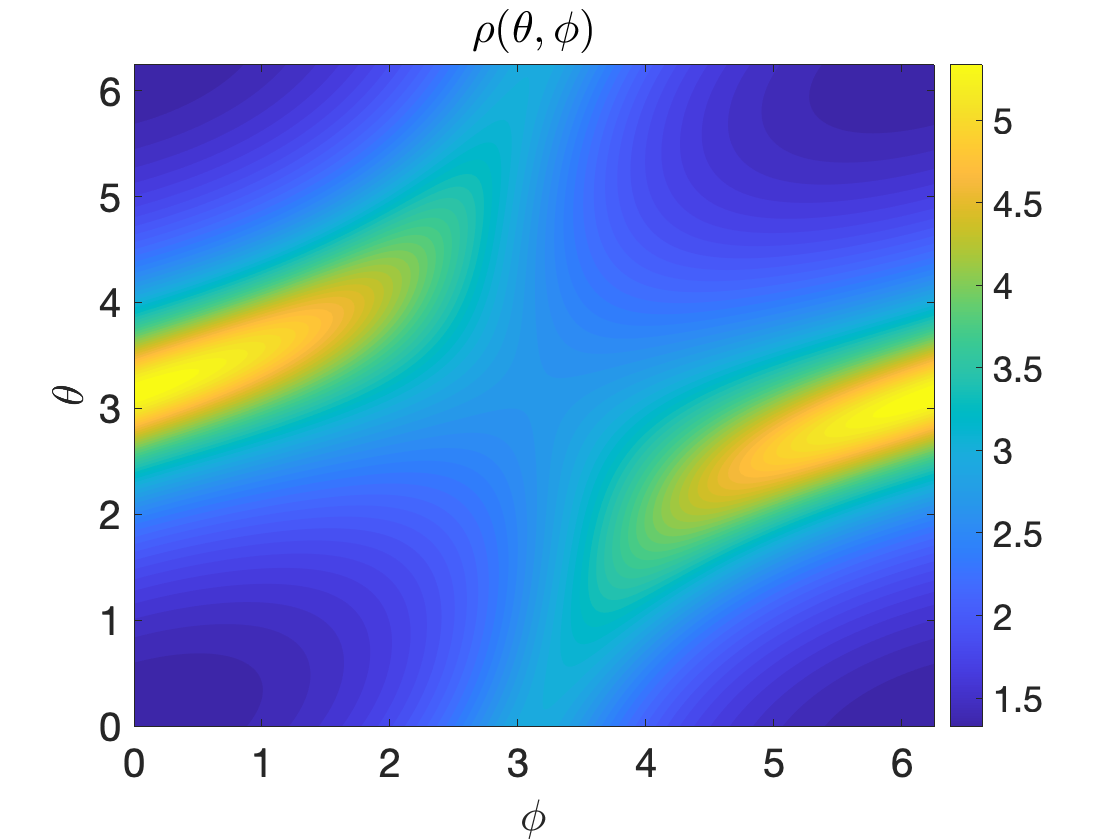}
    \caption{}
    \end{subfigure}
    \begin{subfigure}[b]{0.49\textwidth}
    \includegraphics[width=1.0\textwidth]{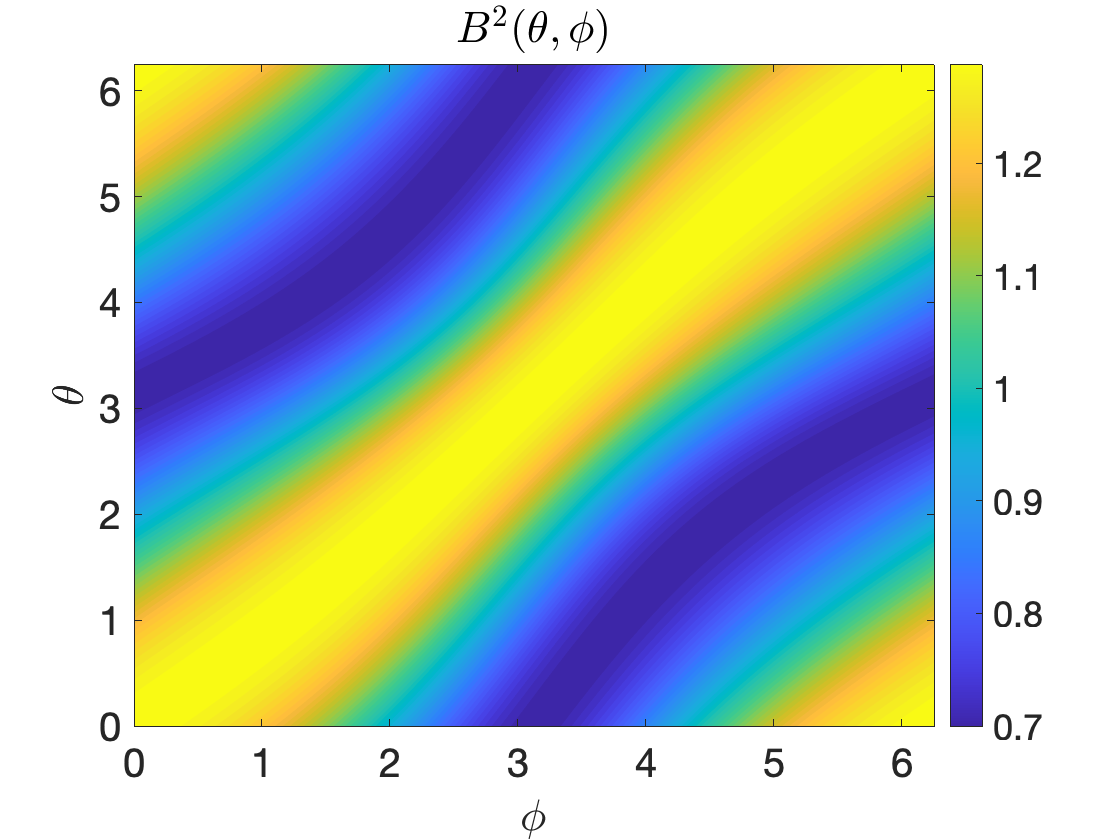}
    \caption{}
    \end{subfigure}
    \begin{subfigure}[b]{0.49\textwidth}
    \includegraphics[width=1.0\textwidth]{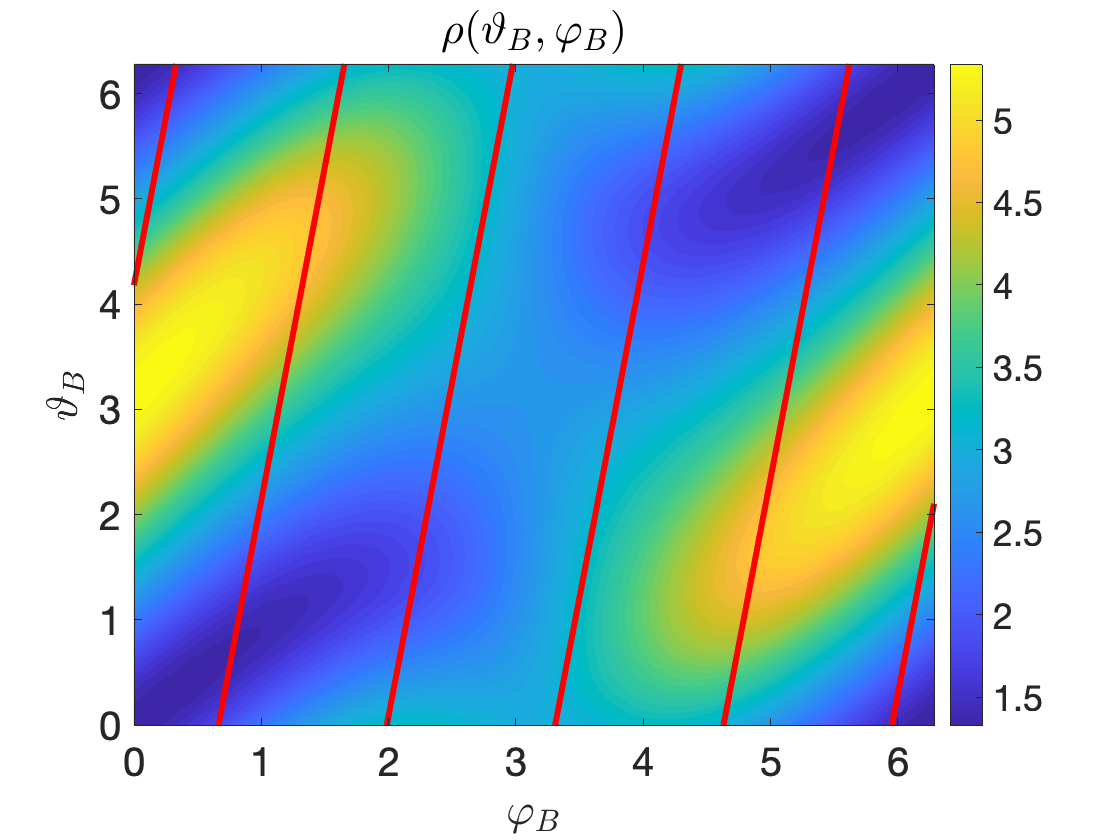}
    \caption{}
    \end{subfigure}
    \begin{subfigure}[b]{0.49\textwidth}
    \includegraphics[width=1.0\textwidth]{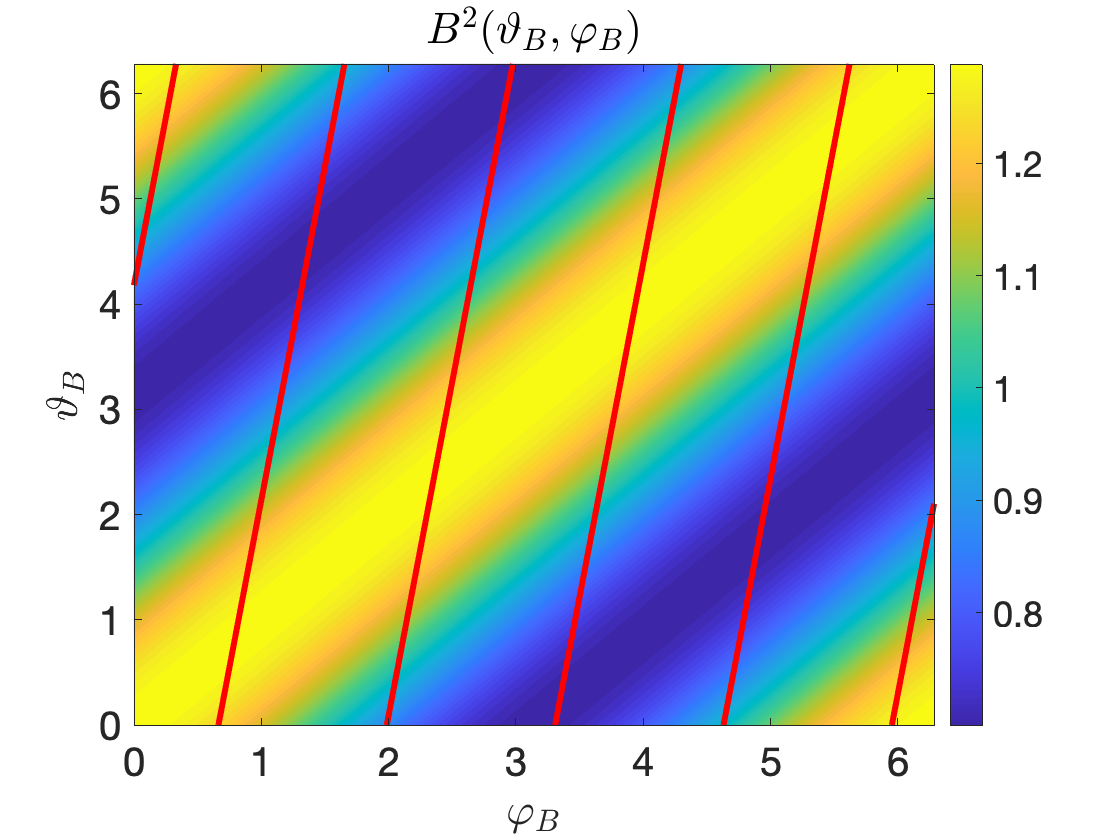}
    \caption{}
    \end{subfigure}
    \caption{Numerical solution for (a) $\rho(\theta,\phi)$, (b) $B^2(\theta,\phi)$, (c) $\rho(\vartheta_B,\varphi_B)$, and (d) $B^2(\vartheta_B,\varphi_B)$ computed imposing the functional form of the field strength given by \eqref{eq:field_strength} with $\epsilon = 0.3$ near the axisymmetric surface given by \eqref{eq:boundary_tok}. The red lines in (c) and (d) indicate field lines.}
    \label{fig:epsilon_0.3}
\end{figure}

\begin{figure}
    \centering
    \begin{subfigure}[b]{0.49\textwidth}
    \includegraphics[width=1.0\textwidth]{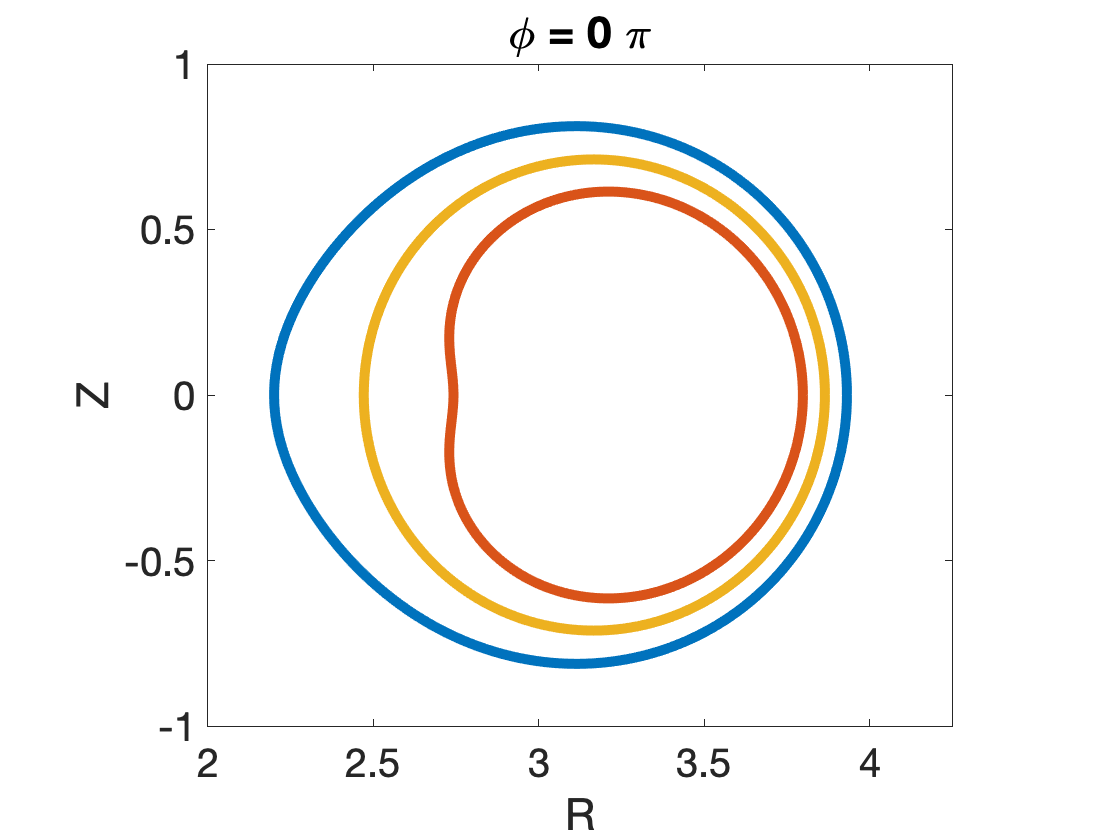}
    \end{subfigure}
    \begin{subfigure}[b]{0.49\textwidth}
    \includegraphics[width=1.0\textwidth]{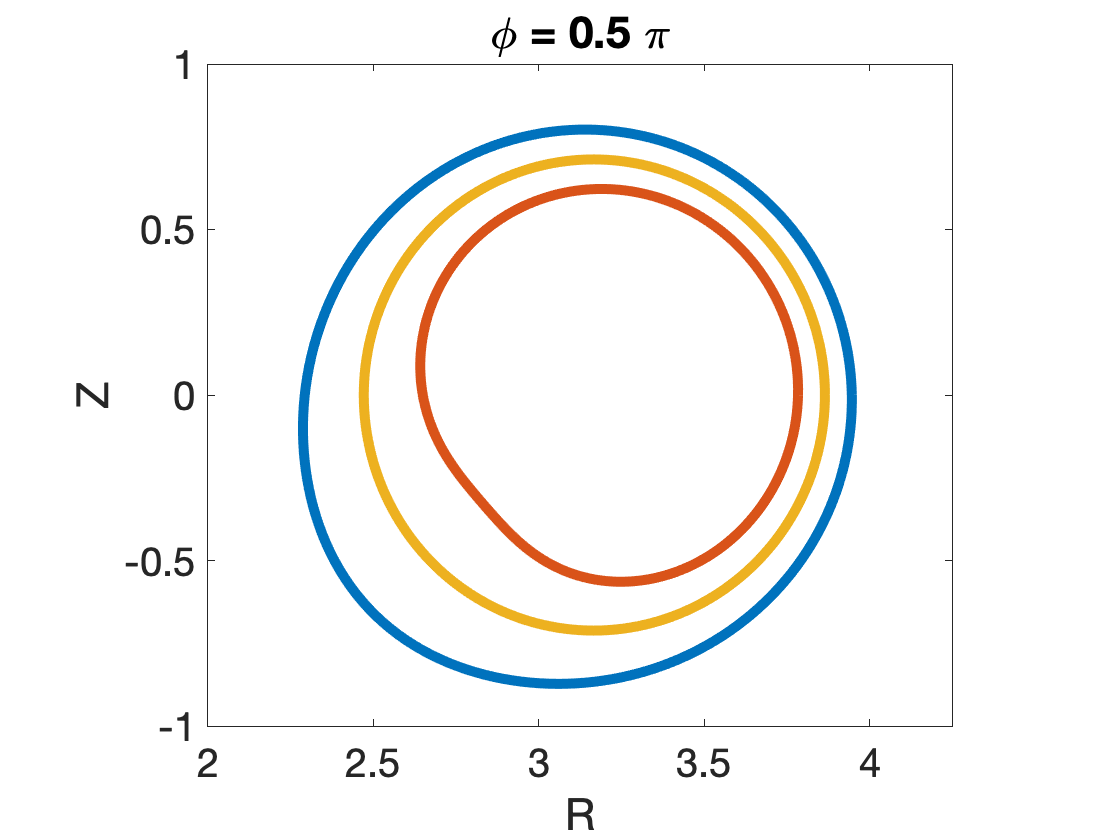}
    \end{subfigure}
    \begin{subfigure}[b]{0.49\textwidth}
    \includegraphics[width=1.0\textwidth]{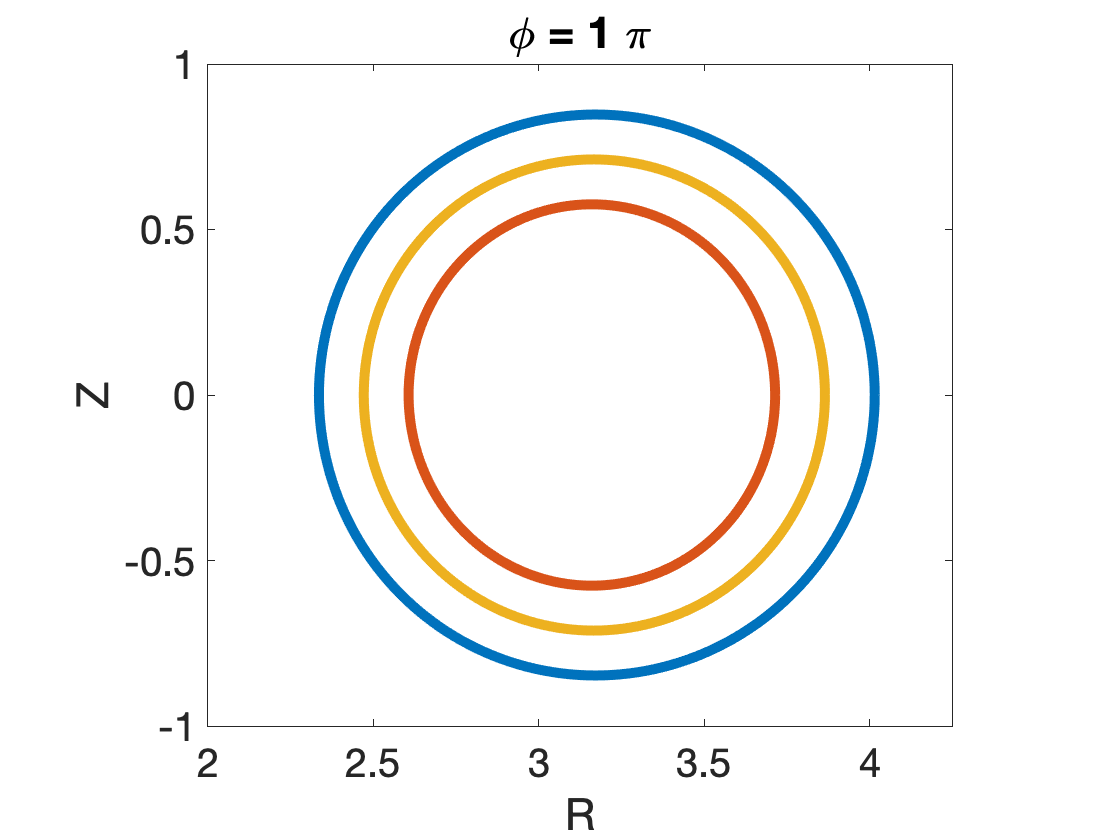}
    \end{subfigure}
    \begin{subfigure}[b]{0.49\textwidth}
    \includegraphics[width=1.0\textwidth]{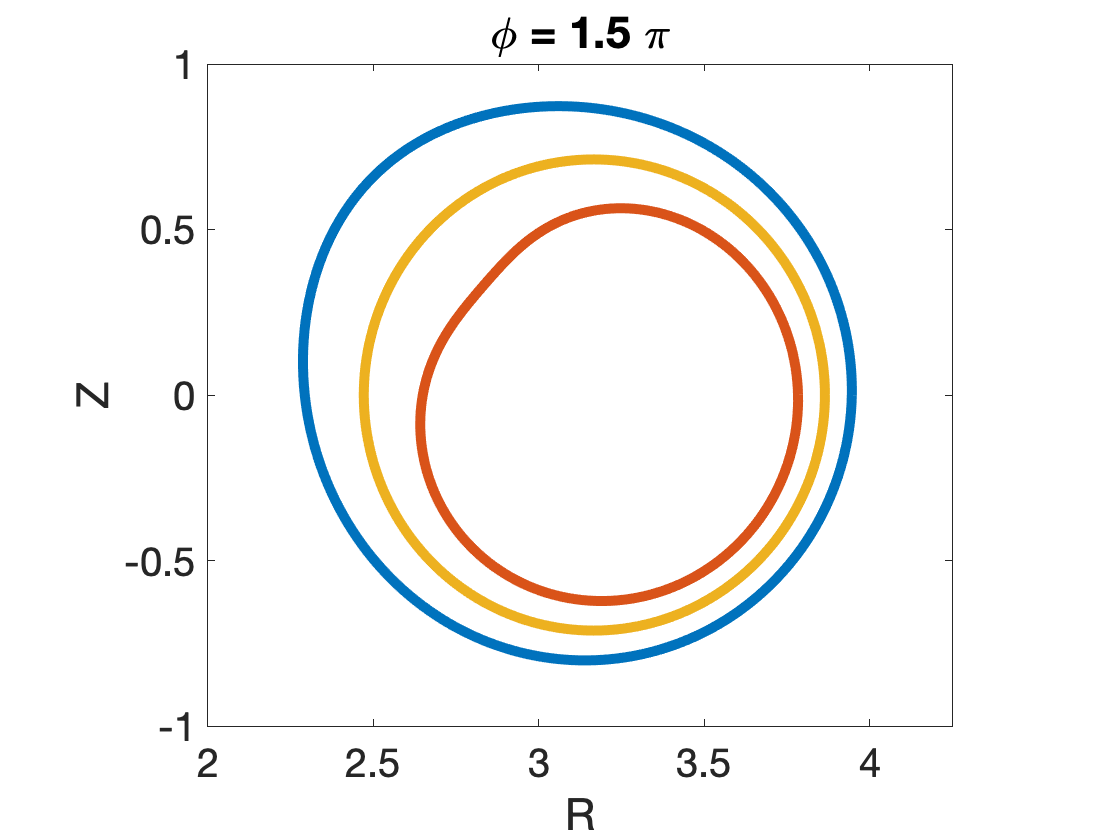}
    \end{subfigure}
    \caption{Using the solution presented in Figure \ref{fig:epsilon_0.3}, we plot the the surface of expansion given by \eqref{eq:boundary_tok} (yellow) and the nearby surfaces (red and blue) given by \eqref{eq:nearby_surfaces}.}
    \label{fig:epsilon_0.3_xc}
\end{figure}

With this method, we are able to obtain a class of helically symmetric solutions near an axisymmetric surface. We prescribe the field strength to be of the form,
\begin{align}
    B^2(\eta) = 1 + \epsilon \cos(\eta),
    \label{eq:field_strength}
\end{align} 
and consider the boundary given by \eqref{eq:boundary_tok}. We present a solution with $\epsilon = 0.3$ in Figures \ref{fig:epsilon_0.3} and \ref{fig:epsilon_0.3_xc}. The nearby surfaces are plotted with a finite value of $\Delta \psi = 0.05$,
\begin{subequations}
\begin{align}
    R^{\pm}(\theta,\phi) &= R(\theta,\phi) \pm \Delta \psi \left(\hat{\textbf{n}} \cdot \hat{\textbf{R}} \right) \rho(\theta,\phi) \\
    Z^{\pm}(\theta,\phi) &= Z(\theta,\phi) \pm \Delta \psi \left(\hat{\textbf{n}} \cdot \hat{\textbf{z}}\right) \rho(\theta,\phi).
\end{align}
\label{eq:nearby_surfaces}
\end{subequations}
We note that the surface perturbations are localized on the inboard size of the torus, with a deformation that results in convexity and concavity in the neighboring surfaces. We will discuss the nature of this trend using the models in Section  \ref{sec:toroidal_solutions}
 shortly. Although the background geometry is axisymmetric, we obtain a helical pattern of field strength. As expected, both the field lines and the contours of the field strength become straight when plotted in the Boozer coordinates $(\vartheta_B,\varphi_B)$. In contrast, $\rho$ does not exhibit a clear symmetry direction when plotted in the Boozer coordinates. However, the contours appear to be straightened due to the dependence of $\rho$ on the quantity $G B^2$, which will be discussed below.
 
\begin{figure}
    \centering
    \begin{subfigure}[b]{0.49\textwidth}
    \includegraphics[width=1.0\textwidth]{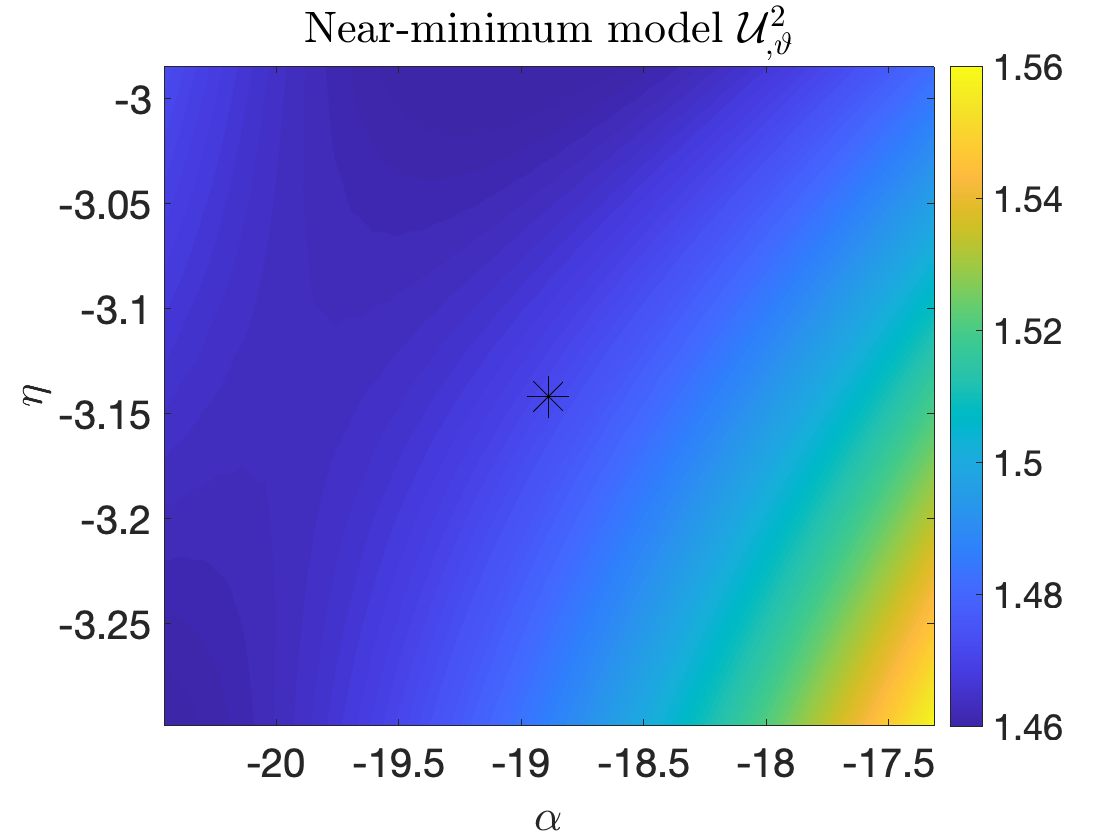}
    \end{subfigure}
    \begin{subfigure}[b]{0.49\textwidth}
    \includegraphics[width=1.0\textwidth]{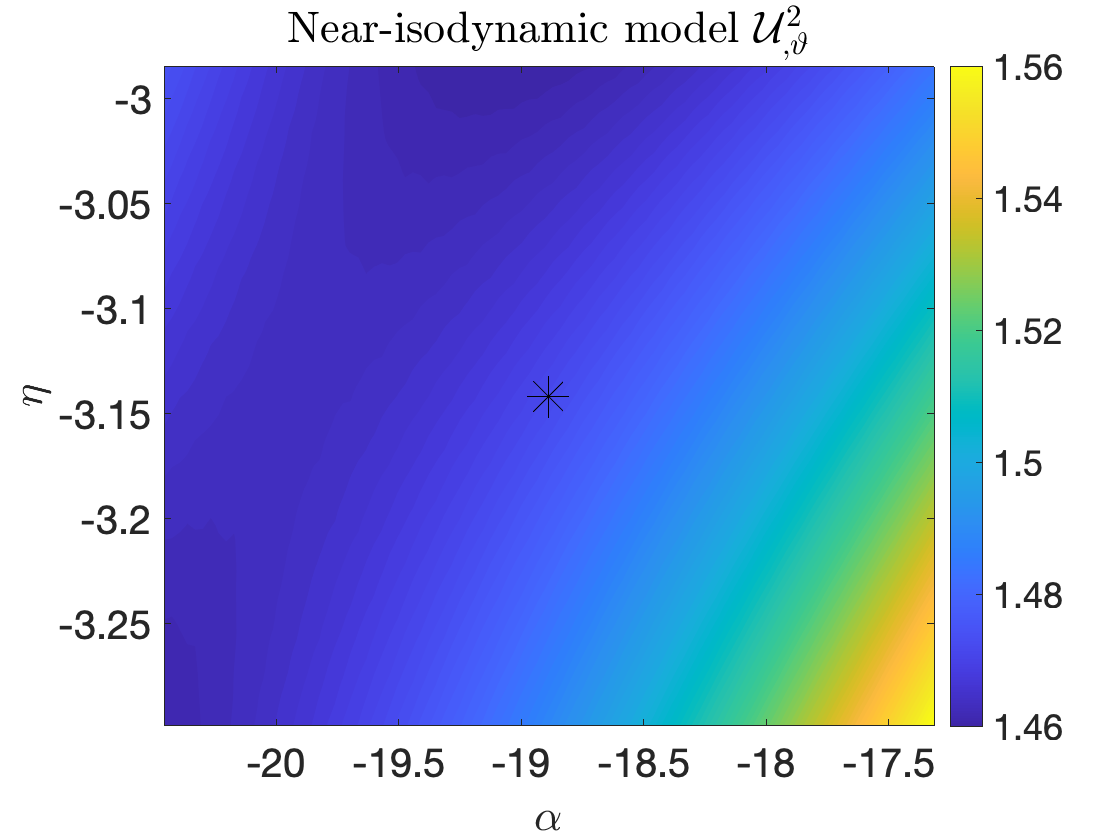}
    \end{subfigure}
    \begin{subfigure}[b]{0.49\textwidth}
    \includegraphics[width=1.0\textwidth]{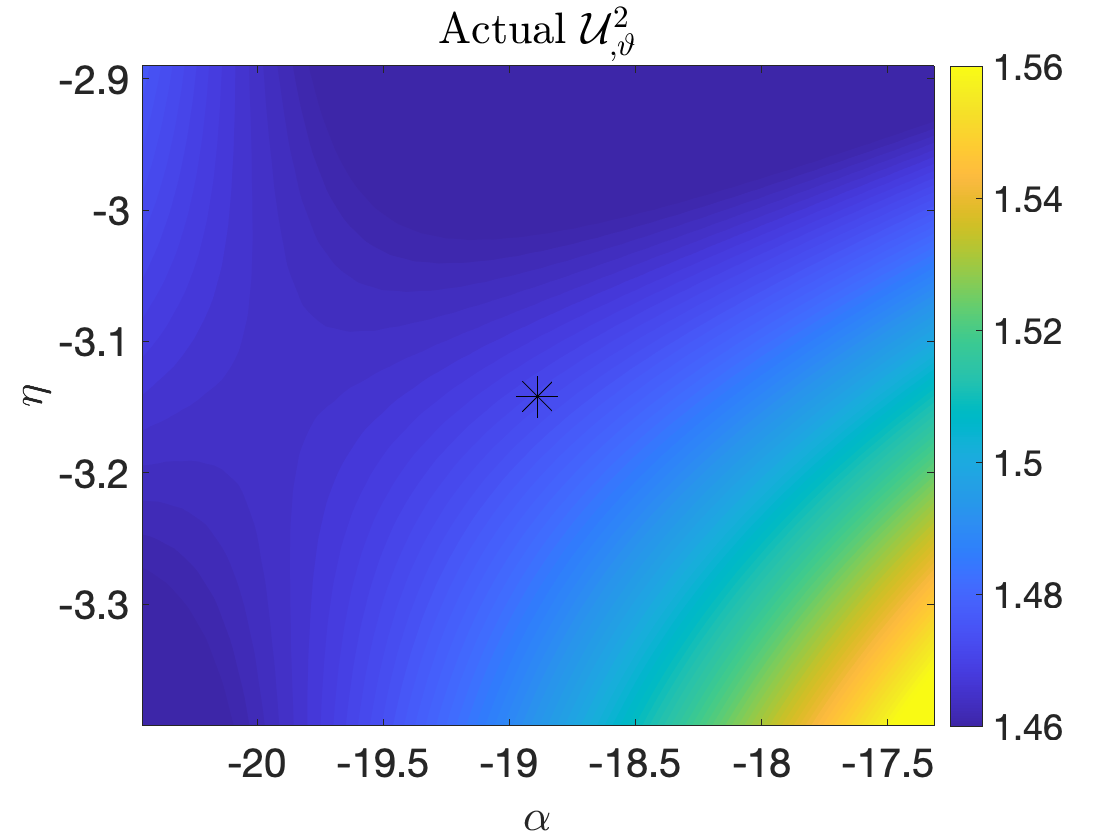}
    \end{subfigure}
    \caption{We compare the solution presented in Figure \ref{fig:epsilon_0.3} with the near-minimum (Section \ref{Near_Bp0}) and near-isodynamic (Section \ref{model_TW_solution}) models. The model is constructed by expanding about a point indicated by the black star.}
    \label{fig:local_expansion}
\end{figure}

 We now compare the models in Section \ref{sec:toroidal_solutions} to this numerical solution. To compare with the model near $B'(\eta) = 0$ in Section \ref{Near_Bp0}, we expand near the minimum of $B(\eta)$ at $\eta = \eta_0 = \pi$ and a chosen value of $\alpha = \alpha_0$. The constant $\Gamma$ is obtained using the value of $\mathcal{U}_{,F \alpha}$ at $(\alpha_0,\eta_0)$. We then note that,
 \begin{align}
     B^2(\delta \mathcal{U}^2) \approx B_0^2 + 2 B_2 B_0 \delta \mathcal{U}^2,
 \end{align}
 so that the exponential in \eqref{eq:c0_near_isodynamic} can be expressed in terms of $B_0$ and $B^2$ rather than $B_2$. The constant $c_0$ is then evaluated using \eqref{eq:c0_near_isodynamic} with the solution for $\mathcal{U}_{,\vartheta}$ and $B^2$ at $(\alpha_0,\eta_0)$. The same expression is then used to model $\mathcal{U}_{,\vartheta}$ in the neighborhood of $(\alpha_0,\eta_0)$, which is denoted by a black star in Figure \ref{fig:local_expansion}. This local model as well as the numerical solution are shown. For the range of $\alpha$ and $\eta$ displayed ($\alpha \in [\alpha_0-\pi/2,\alpha_0+\pi/2]$ and $\eta \in [\eta_0 - \pi/20,\eta_0+\pi/20]$), the averaged relative error,
 \begin{align}
     \text{error} = \frac{\int d \alpha \int d \eta \, \left|  \mathcal{U}_{,\vartheta}^2 - \left(\mathcal{U}_{,\vartheta}^{\text{model}}\right)^2 \right| }{\int d \alpha \int d \eta \, \mathcal{U}_{,\vartheta}^2},
 \end{align}
is $1.8\times10^{-3}$. 

We also compare with the near-isodynamic limit \eqref{model_TW_solution}. We expand about the same point $(\alpha_0,\eta_0)$ and again fix $\Gamma$ to be the value of $\mathcal{U}_{,F\alpha}$ at this point. The constant $c_0$ is then fixed using the value of $\mathcal{U}_{,\vartheta}$ at this point, and \eqref{model_TW_solution} is used to model $\mathcal{U}_{,\vartheta}$ in the neighborhood of $(\alpha_0,\eta_0)$. The result is shown in Figure \ref{fig:local_expansion}. The averaged error between the numerical solution and the model is $2.0\times 10^{-3}$. 

\begin{figure}
    \centering
    \begin{subfigure}[b]{0.49\textwidth}
     \includegraphics[width=1.0\textwidth]{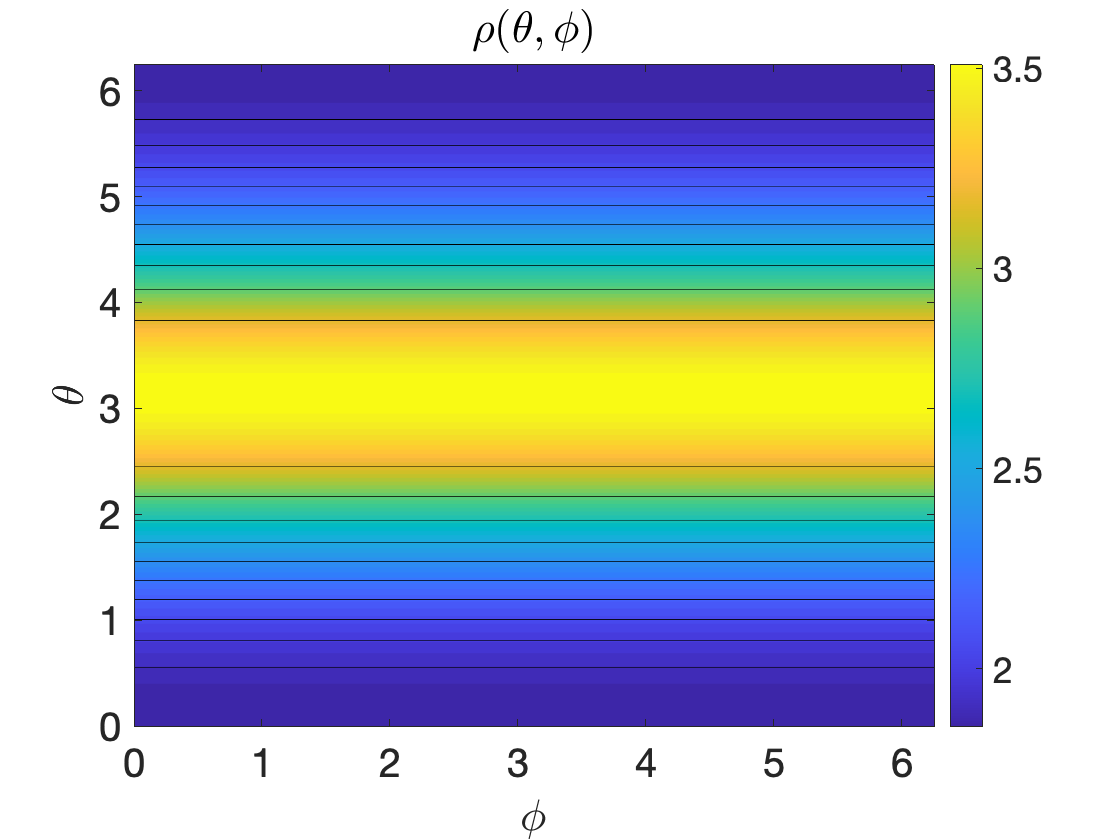}
    \caption{$\epsilon = 0$} 
    \end{subfigure}
    \begin{subfigure}[b]{0.49\textwidth}
    \includegraphics[width=1.0\textwidth]{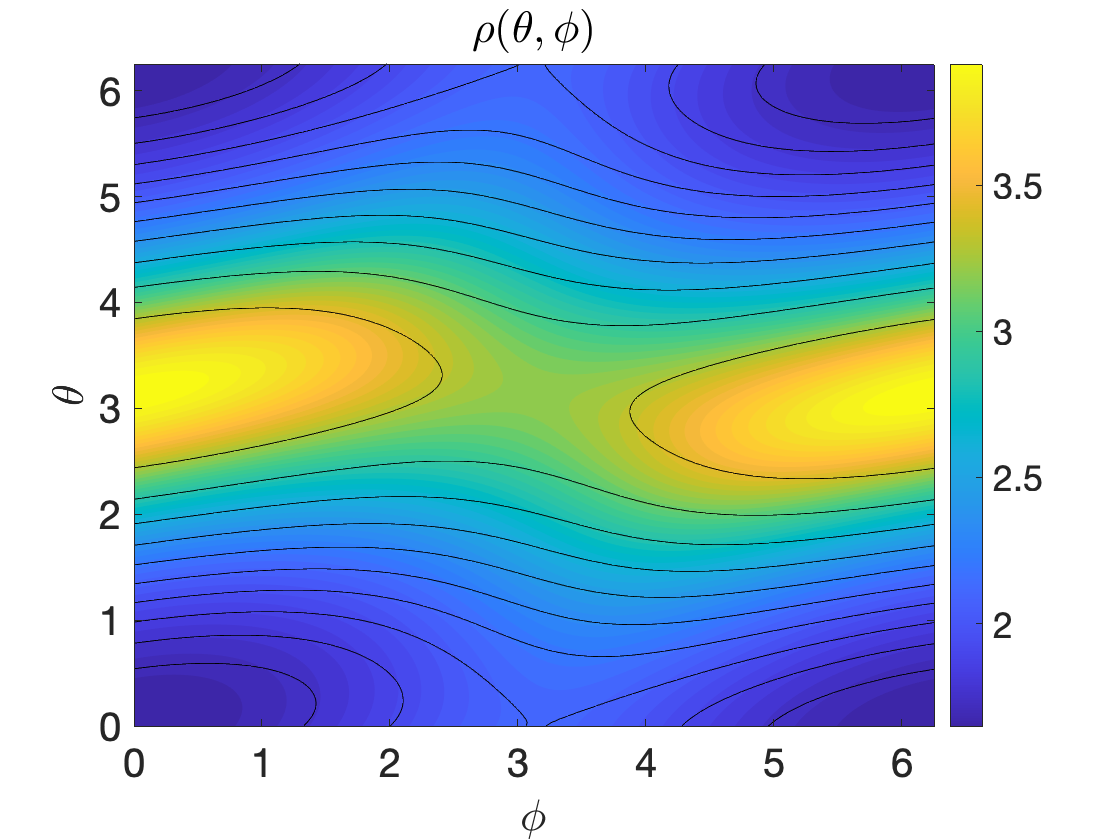}
    \caption{$\epsilon = 0.1$}
    \end{subfigure}
    \begin{subfigure}[b]{0.49\textwidth}
    \includegraphics[width=1.0\textwidth]{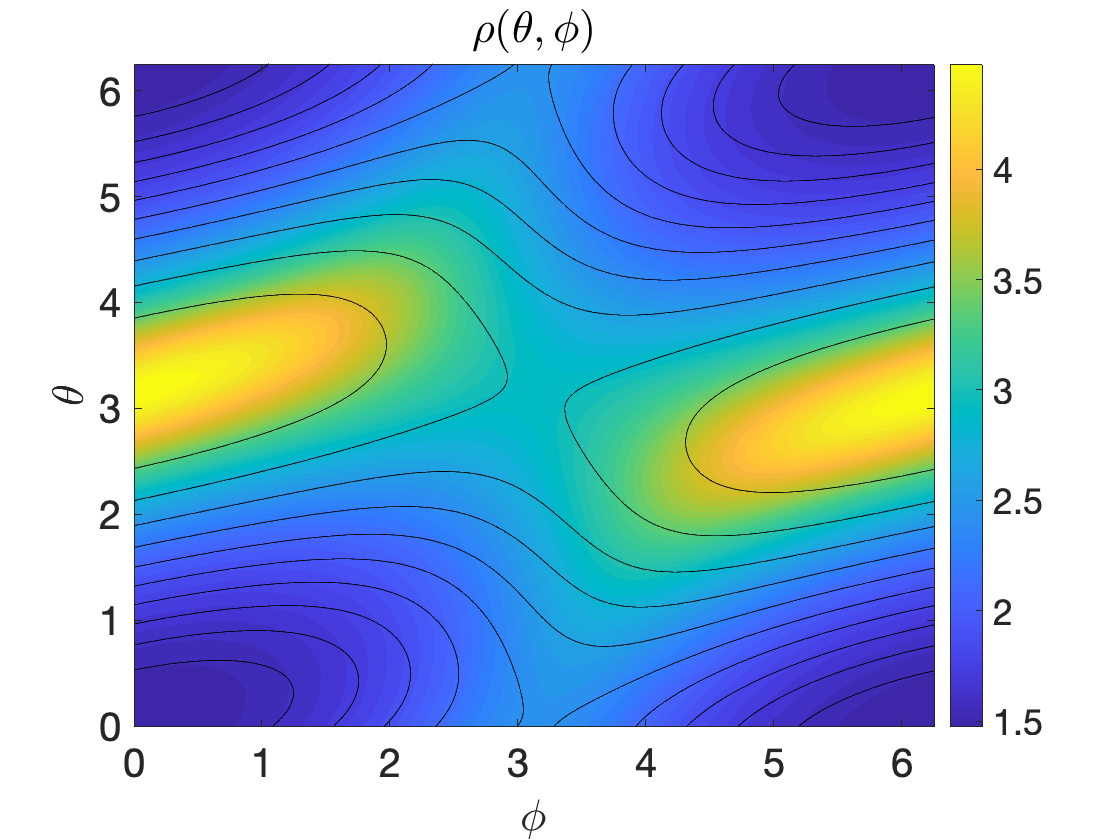}
    \caption{$\epsilon = 0.2$}
    \end{subfigure}
    \begin{subfigure}[b]{0.49\textwidth}
    \includegraphics[width=1.0\textwidth]{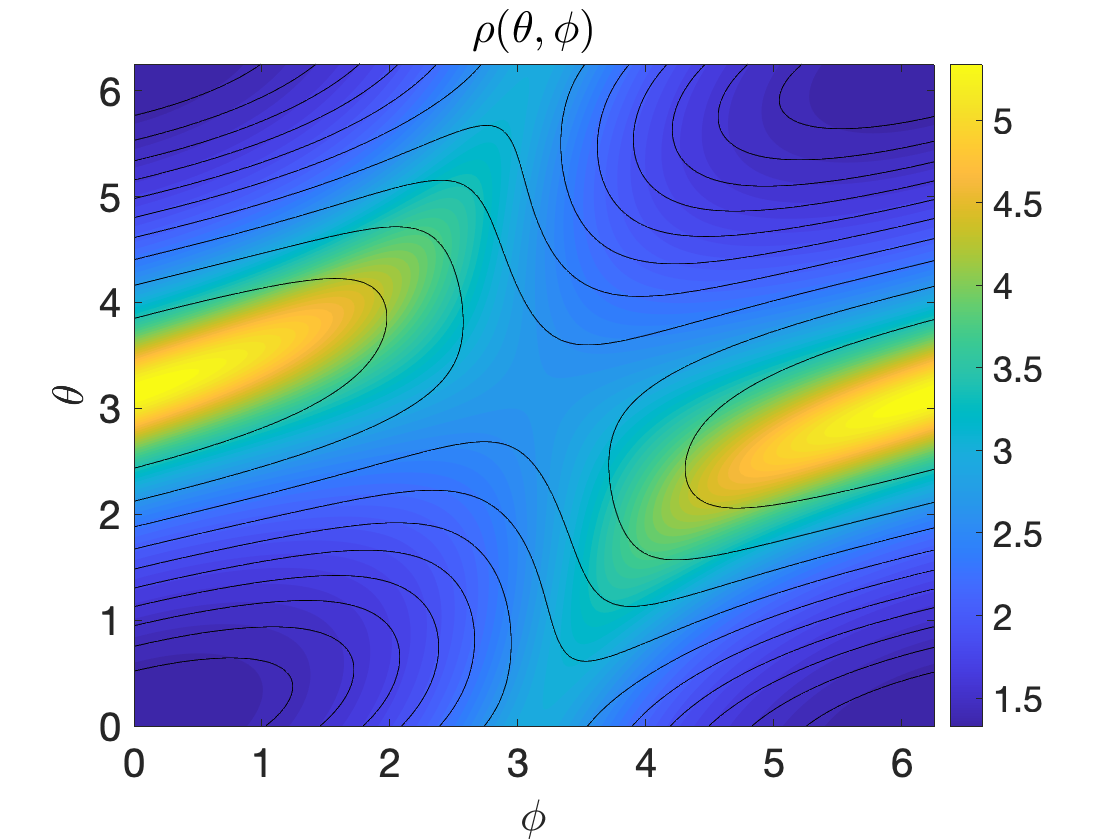}
    \caption{$\epsilon = 0.3$}
    \end{subfigure}
    \begin{subfigure}[b]{0.49\textwidth}
    \includegraphics[width=1.0\textwidth]{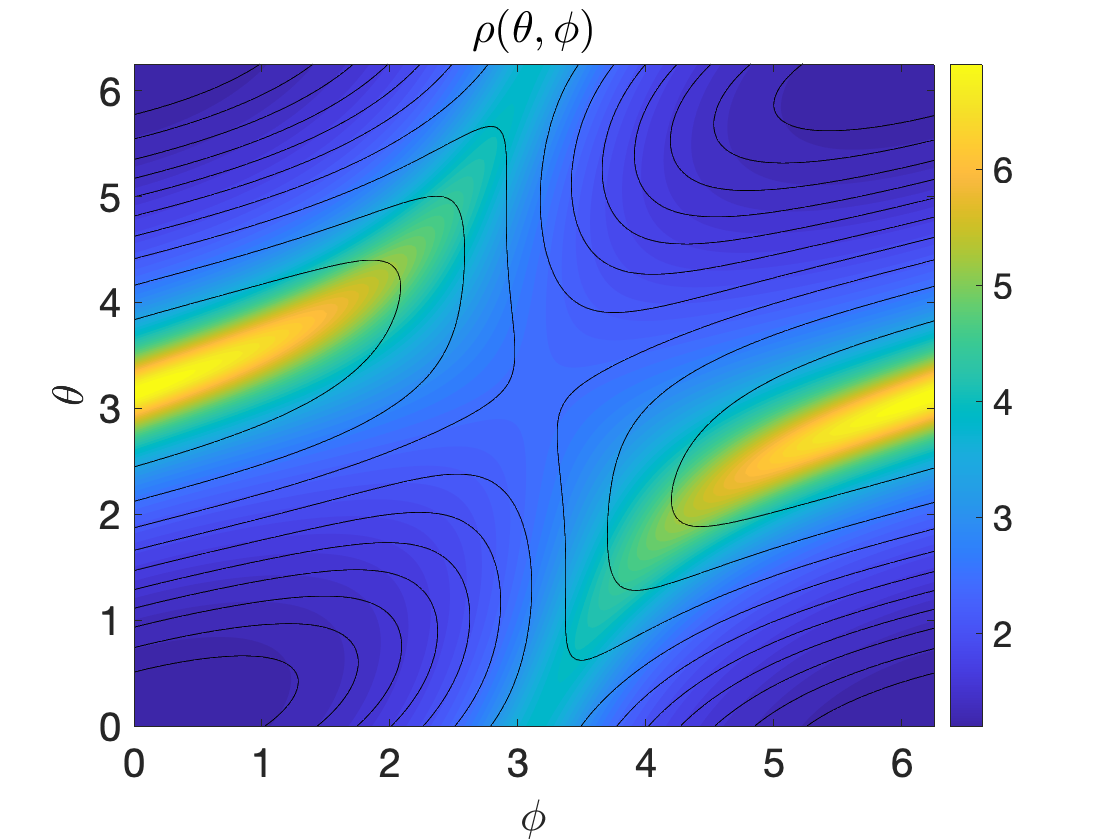}
    \caption{$\epsilon = 0.4$}
    \end{subfigure}
    \begin{subfigure}[b]{0.49\textwidth}
    \includegraphics[width=1.0\textwidth]{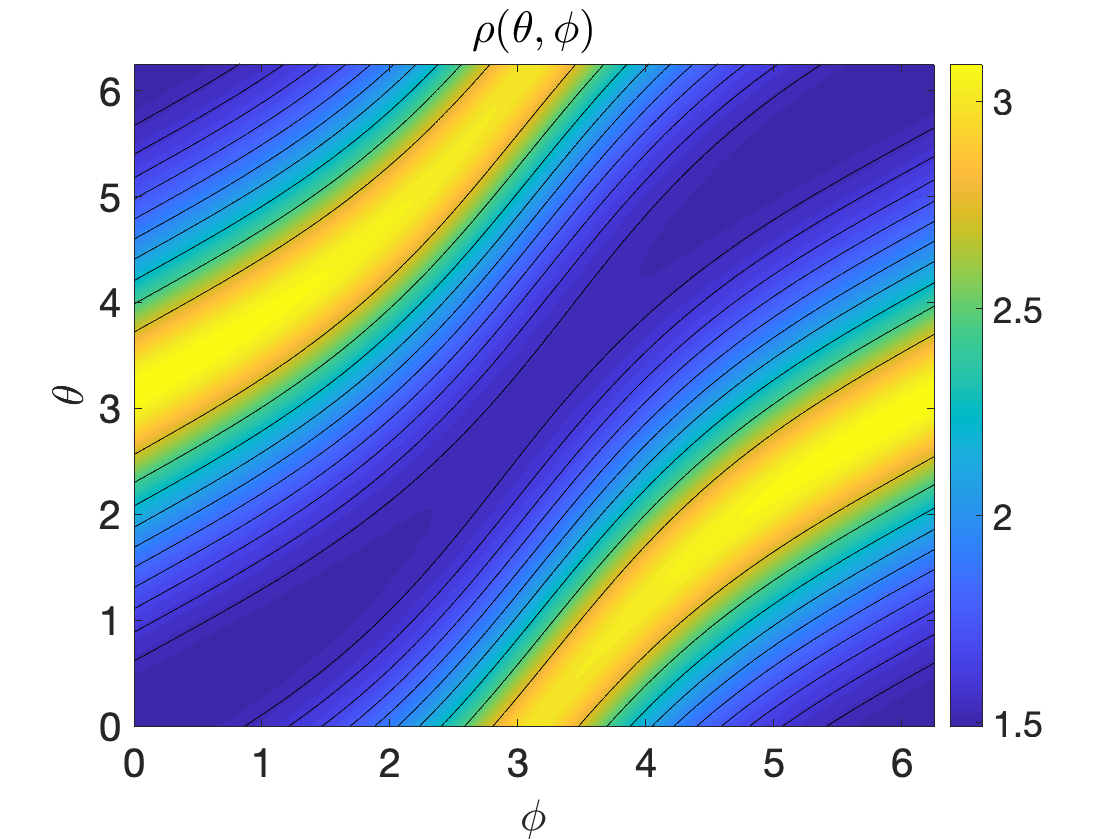}
    \caption{$\epsilon = 0.3$, $R_0 = 100$}
    \label{fig:R0_100}
    \end{subfigure}
    \caption{We present numerical solutions using the functional form of the field strength given by \eqref{eq:field_strength} with the displayed value of $\epsilon$ near the axisymmetric surface given by \eqref{eq:boundary_tok}. The colorscale indicates the magnitude of $\rho(\theta,\phi)$, with the overlaid black lines indicating the contours of $GB^2$. In (f), the boundary given by \eqref{eq:boundary_tok} is modified such that the major radius is $R_0 = 100$.}
    \label{fig:rho_gB2}
\end{figure}

While the near-minimum and near-isodynamic models are only valid in the neighborhood of the minimum of the field strength, the models allow us to glean insight into the global nature of the solution. In particular, we note that the spatial dependence of $\mathcal{U}_{,\vartheta}$ and $\rho$ is largely through the combination $G B^2$. In Figure \ref{fig:rho_gB2} we present the solution for $\rho(\theta,\phi)$ for several values of $\epsilon$. We note that as $\epsilon$ is increased, the perturbation of the nearby surfaces becomes highly localized, introducing sharp helical structures in $\rho$. For smaller values of $\epsilon$, the maximum of $\rho$ becomes localized on the inboard side where $G$ is minimized. In the limit that $\epsilon\rightarrow 0$, $\rho(\theta,\phi)$ becomes axisymmetric. The black contours overlaid on the colorscale plots in Figure \ref{fig:rho_gB2} are contours of $G B^2$, with the minimum of $GB^2$ coinciding with the maximum of $\rho$. While the contours of $GB^2$ do not exactly coincide with the contours of $\rho$ for $\epsilon>0$, we note that it captures much of the structure of the numerical solution. The results we present here are consistent with features of near-axisymmetric QA solutions \citep{Plunk2018,Plunk2020}, for which the non-axisymmetric perturbations to the field were localized to the inboard side. In Figure \ref{fig:R0_100}, the major radius of the boundary given by \eqref{eq:boundary_tok} is increased to $R_0 = 100$. In this case, $G(\vartheta)$ approaches a constant such that the perturbation to the surfaces is no longer localized to the inboard side. The large aspect ratio toroidal solution then approaches the traveling wave (helical) solutions of the slab and cylindrical geometry.   

\FloatBarrier
\section{Discussion}
\label{sec:discussion}

In this work, we have introduced the formalism for obtaining vacuum magnetic fields with exact quasisymmetry on a flux surface. Quasisymmetry is enforced by introducing an unknown, $\eta$, such that $B(\eta)$. In solving the relevant equations at the lowest order, the free functions are the shape of the flux surface, the functional form of $B(\eta)$, and the desired helicity of the quasisymmetry. This is somewhat analogous to the free functions that arise in the axis expansion equations, for which the axis shape is specified in addition to certain harmonics of the field strength. While the shape of the axis determines the helicity of the quasisymmetry, the helicity can be specified in the surface-expansion problem.

As an initial step, we have made several simplifying assumptions in this work. We have focused on the vacuum equations, but the formalism can be easily extended to model force-free or MHD equilibrium fields following \cite{Weitzner2016}. We have also focused on solutions near an axisymmetric surface. While the surface itself is axisymmetric, non-axisymmetry is introduced in the nearby surfaces and is determined self-consistently as part of the solution. While the introduction of non-axisymmetry in the background geometry complicates the metric tensor (Appendix \ref{sec:torus}), the formalism is not limited to axisymmetry. 

Interestingly, we have shown that the resulting equation for $\eta$ is generally parabolic (Appendix \ref{app:parbolic}). This in contrast to the generalized Grad-Shafranov equation for the flux function in quasisymmetry \citep{Burby2020}, which is elliptic in nature. We leave further analysis of the properties of this nonlinear parabolic PDE to future work. Whether the PDE for $\eta$ is of parabolic type for force-free or MHD equilibria remains to be shown.

The solutions presented in this work have been obtained under the traveling wave ansatz such that the field strength is helical. For this reason, we were able to converge to quasi-helical solutions in general toroidal geometry by initializing with a traveling wave solution. In future work, we aim to explore quasi-axisymmetric and other possible classes of solutions within this formalism.

Although here we have taken a local equilibrium approach, which does not guarantee existence within a global equilibrium, we can extend the formalism to look for global traveling wave solutions in cylindrical and slab geometries. This is a special class of solutions which, if it exists, should satisfy the overdetermined system of equations and will be reported in subsequent work. 






\noindent{\bf Acknowledgements}
The authors would like to thank J. McFadden, E. Kim, G. Plunck, G. Roberg-Clark, M. Landreman, R. Jorge, and A. Cerfon for helpful suggestions. This research was partly
funded by the US DOE grant no. DEFG02-86ER53223.

\appendix
\section{Various geometries \label{Appendix_geometries}}
We shall now give the details of the various geometries that naturally arise in stellarator theory. The simplest geometry is the slab geometry (Section \ref{sec:slab}) with doubly-periodic boundary conditions. In Section \ref{sec:torus} we discuss toroidal geometry, which is equivalent to a periodic cylinder in the infinite aspect ratio limit. 


\subsection{Slab geometry}
\label{sec:slab}
We consider a Cartesian coordinate system $(x,y,z)$. Given surfaces labeled by the toroidal flux $\psi$, we perform a transformation to the coordinate system $(\psi,y,z)$ where $y$ and $z$ are $2\pi$ periodic. Given the derivatives of the position vector,
\begin{subequations}
\begin{align}
    \textbf{r}_{,\psi} &= x_{,\psi}(\psi,y,z) \hat{\textbf{x}} \\
    \textbf{r}_{,y} &= x_{,y}(\psi,y,z) \hat{\textbf{x}} + \hat{\textbf{y}} \\
    \textbf{r}_{,z} &= x_{,z}(\psi,y,z) \hat{\textbf{x}} + \hat{\textbf{z}},
\end{align}
\end{subequations}
we obtain the following metric coefficients under the assumption of axisymmetry ($x_{,z} = 0$),
\begin{subequations}
\begin{align}
    g_{\psi y} &= x_{,\psi} x_{,y} \\
    g_{yy} &= 1 + x_{,y}^2 \\
    g_{zz} &= 1  \\
    g_{\psi z} &= g_{yz} = 0,
\end{align}
\end{subequations}
and the full flux coordinate Jacobian is given by,
\begin{align}
\sqrt{g} = \textbf{r}_{,\psi} \cdot \textbf{r}_{,y} \times \textbf{r}_{,z} &= x_{,\psi}.
\end{align}


Given these geometric quantities, the vacuum equations in axisymmetric slab geometry take the form
\begin{subequations}
\begin{align}
x_{,\psi}\begin{pmatrix}
\Phi_{,y}\\ \Phi_{,z}
\end{pmatrix}=
\begin{pmatrix}
\cE \quad  \cF\\ 
\cF \quad  \cG
\end{pmatrix}
\begin{pmatrix}
-\alpha_{,z}\\ \alpha_{,y}
\end{pmatrix} \\
\cE=  1 + x_{,y}^2 , \quad \cF = 0,\quad \cG= 1\\
\Phi_{,\psi}=x_{,\psi}\BD x=\{\alpha,x\}_{(y,z)} \label{eq:Phi_psi}\\
B^2= \BD \Phi = \frac{1}{x_{,\psi}}\{\alpha,\Phi\}_{(y,z)}. 
\end{align}
\label{QS_vacuum_slab}
\end{subequations}
Note that in this coordinate system, $\Phi_{,\psi}$ does not vanish as it did for the local coordinate system described by \eqref{eq:local_coordinates}. Thus \eqref{eq:Phi_psi} can be used to determine the first-order correction to the scalar potential, but this constraint will not be included in the lowest-order system. Defining new variables 
\begin{align}
 \rho=\frac{x_{,\psi}}{\sqrt{1+x'^2}}, \quad \Theta= \int\: dy\sqrt{1+x'^2} ,\quad \phi=z  
\end{align}
the in-surface equations can be cast in the following form
\begin{subequations}
\begin{align}
\rho \begin{pmatrix}
\Phi_{,\vartheta}\\ \Phi_{,\phi}
\end{pmatrix}=
\begin{pmatrix}
-\alpha_{,\phi}\\ \alpha_{,\Theta}
\end{pmatrix} \\
B^2=  \frac{1}{\rho}\{\alpha,\Phi\}_{(\Theta,\phi)}.
\end{align}
\end{subequations}
Thus we recover the system obtained in the main text \eqref{eq:QS_system1} with $g_{\phi\phi} = 1$.

\subsection{Cylindrical and toroidal geometry}
\label{sec:torus}

We consider the coordinate system $(\psi,\theta,\phi)$ where $\theta$ and $\phi$ are 2$\pi$ periodic. We map from flux coordinate space to real space using the radius function, $r(\theta,\phi)$, from a fixed coordinate axis, $\{R_0(\phi),Z_0(\phi)\}$ where
\begin{subequations}
\begin{align}
R(\psi,\theta,\phi) &= R_0(\phi) + r(\psi,\theta,\phi)\cos(\theta) \\
Z(\psi,\theta,\phi) &= Z_0(\phi) + r(\psi,\theta,\phi)\sin(\theta).
\end{align}
\end{subequations}
The derivatives of the position vector are given by,
\begin{subequations}
\begin{align}
    \textbf{r}_{,\psi} &= r_{,\psi} \left(\cos(\theta)\hat{\textbf{R}} + \sin(\theta)\hat{\textbf{z}}\right) \\
    \textbf{r}_{,\theta} &=  r_{,\theta}\left(\cos(\theta)\hat{\textbf{R}} + \sin(\theta)\hat{\textbf{z}}\right) + r \left(-\sin(\theta)\hat{\textbf{R}} + \cos(\theta)\hat{\textbf{z}} \right) \\
    \textbf{r}_{,\phi} &=  -\left(R_0 + r\cos(\theta)\right)\hat{\bm{\phi}},
\end{align}
\end{subequations}
under the assumption of axisymmetry ($R_0'(\phi) = Z_0'(\phi) = r_{,\phi} = 0$).

The metric elements are then given by,
\begin{subequations}
\begin{align}
    g_{\psi\theta} &= r_{,\psi} r_{,\theta} \\
    g_{\theta \theta} &= r_{,\theta}^2 + r^2 \\
    g_{\phi \phi} &= \left(R_0 + r\cos(\theta)\right)^2 \\
    g_{\psi \phi} &= g_{\theta \phi} = 0,
\end{align}
\end{subequations}
and the Jacobian is,
\begin{align}
    \sqrt{g} = \textbf{r}_{,\psi} \cdot \textbf{r}_{,\theta} \times \textbf{r}_{,\phi} = r_{,\psi} r^2 .
\end{align}
Under the cylindrical approximation, we take $z = R_0 \phi$ to be the axial direction and note that $\textbf{r}_{,z} = -\hat{\textbf{z}}$ under the assumption that $r/R_0 \ll 1$. Thus $g_{zz} = 1$.

\section{Parabolic nature of $\eta$ equation}
\label{app:parbolic}

In this Appendix, we show that the PDE for $\eta$ remains parabolic for a general metric. We begin with the vacuum equations and quasisymmetry constraint in general geometry,
\begin{subequations}
\begin{align}
\frac{\rho F}{\mathcal{J}\left(1 + F^2 \rho^2 \right)}\begin{pmatrix}
\quad g_{\phi\phi} \quad \quad \quad  -g_{\theta\phi} - F \rho \mathcal{J}\\ 
\: -g_{\theta\phi} + F\rho \mathcal{J} \quad  \quad\quad  g_{\theta\theta}\quad \quad
\end{pmatrix}
\begin{pmatrix}
\eta_{,\theta}  \\ \eta_{,\phi}
\end{pmatrix} 
&= \begin{pmatrix}
-F\alpha_{,\phi} \\ +F\alpha_{,\theta}
\end{pmatrix}
\label{gen_geom_eta}
\end{align}
\begin{align}
    \rho\mathcal{J} B^2(\eta) &= \eta_{,\phi}\alpha_{,\theta}-\eta_{,\theta}\alpha_{,\phi}.
\end{align}
\end{subequations}
Combining these two, we obtain an equation for $\rho F$,
\begin{align}
     1 + F^2 \rho^2 &=\frac{\eta_{,\phi}^2 g_{\theta\theta} + \eta_{,\theta}^2 g_{\phi\phi} - 2 g_{\theta \phi} \eta_{,\theta} \eta_{,\phi}}{\mathcal{J}^2 B^2(\eta)}.
     \label{eq:general_rhoF}
\end{align}
Now equating mixed partial derivatives of $F\alpha$, we obtain,
\begin{multline}
    \partder{}{\theta} \left[\frac{\rho F}{\mathcal{J}\left(1 + F^2 \rho^2 \right)}\left(g_{\phi\phi}\eta_{,\theta} - \eta_{,\phi}\left(g_{\theta\phi} + F \rho \mathcal{J} \right) \right) \right] \\
    + \partder{}{\phi} \left[\frac{\rho F}{\mathcal{J}\left(1 + F^2 \rho^2 \right)} \left(\eta_{,\theta}\left(-g_{\theta\phi} + F \rho \mathcal{J} \right) + \eta_{,\phi} g_{\theta\theta} \right) \right] = 0.
    \label{eq:parabolic_pde_general}
\end{multline}
After using the expression for $\rho F$ \eqref{eq:general_rhoF}, we obtain a single PDE for $\eta$ which takes the form,
\begin{align}
    A \partial_{\theta}^2 \eta + 2 B \partial_{\theta}\partial_{\phi}\eta + C \partial_{\phi}^2\eta + f \left(\eta,\theta,\phi,\eta_{,\theta}, \eta_{,\phi}\right) = 0,
\end{align}
with coefficients given by,
\begin{subequations}
\begin{align}
    A &= \left[-( g_{\theta \phi} + F \mathcal{J} \rho) \eta_{,\phi} + g_{\phi\phi} \eta_{,\theta} \right]^2 \\
    C &= \left[g_{\theta\theta} \eta_{,\phi} - \eta_{,\theta} ( g_{\theta \phi} - F \mathcal{J} \rho) \right]^2 \\
  B  &=  \left[g_{\theta\theta} \eta_{,\phi} - \eta_{,\theta} ( g_{\theta \phi} - F \mathcal{J} \rho) \right]\left[-( g_{\theta \phi} + F \mathcal{J} \rho) \eta_{,\phi} + g_{\phi\phi} \eta_{,\theta} \right].
\end{align}
\label{ABC_coeffs}
\end{subequations}
We see that this form is consistent with \eqref{eta_eqn} in the limit that $g_{\theta \phi} \rightarrow 0$. Therefore, the PDE for $\eta$ remains parabolic in general geometry. From the coefficients \eqref{ABC_coeffs} and \eqref{gen_geom_eta}, one can see that the repeated characteristics are given by constant $F\alpha$ curves.

\bibliographystyle{jpp}
\bibliography{plasmalit}

\begin{thebibliography}{40}
\expandafter\ifx\csname natexlab\endcsname\relax\def\natexlab#1{#1}\fi
\def\au#1{#1} \def\ed#1{#1} \def\yr#1{#1}\def\at#1{#1}\def\jt#1{\textit{#1}}
  \def\bt#1{#1}\def\bvol#1{\textbf{#1}} \def\vol#1{#1} \def\pg#1{#1}
  \def\publ#1{#1}\def\arxiv#1{#1}\def\org#1{#1}\def\st#1{\textit{#1}}

\bibitem[Bers(1953)]{Bers1953}
{\sc \au{Bers, Lipman}} \yr{1953} {\em Theory of Pseudo-analytic Functions\/}.
  \publ{New York University. Institute for Mathematics and Mechanics}.

\bibitem[Boozer(1983)]{Boozer1983}
{\sc \au{Boozer, Allen~H}} \yr{1983}  \at{Transport and isomorphic equilibria}.
   \jt{The Physics of Fluids}  \bvol{26}~(2),  \pg{496}.

\bibitem[Boozer(2002)]{Boozer2002}
{\sc \au{Boozer, Allen~H}} \yr{2002}  \at{Local equilibrium of nonrotating
  plasmas}.  \jt{Physics of Plasmas}  \bvol{9}~(9),  \pg{3762--3766}.

\bibitem[Boozer(2019{\natexlab{{\em a\/}}})]{Boozer2019b}
{\sc \au{Boozer, Allen~H}} \yr{2019{\natexlab{{\em a\/}}}}  \at{Curl-free
  magnetic fields for stellarator optimization}.  \jt{Physics of Plasmas}
  \bvol{26}~(10),  \pg{102504}.

\bibitem[Boozer(2019{\natexlab{{\em b\/}}})]{Boozer2019}
{\sc \au{Boozer, Allen~H}} \yr{2019{\natexlab{{\em b\/}}}}  \at{Stellarators as
  a fast path to fusion energy}.  \jt{arXiv preprint arXiv:1912.06289} .

\bibitem[Burby {\em et~al.\/}(2020)Burby, Kallinikos \& MacKay]{Burby2020}
{\sc \au{Burby, JW}, \au{Kallinikos, N} \& \au{MacKay, RS}} \yr{2020}  \at{Some
  mathematics for quasi-symmetry}.  \jt{Journal of Mathematical Physics}
  \bvol{61}~(9),  \pg{093503}.

\bibitem[Candy \& Belli(2015)]{Candy2015}
{\sc \au{Candy, J} \& \au{Belli, Emily~A}} \yr{2015}  \at{Non-axisymmetric
  local magnetostatic equilibrium}.  \jt{Journal of Plasma Physics}
  \bvol{81}~(3).

\bibitem[Constantin {\em et~al.\/}(2020)Constantin, Drivas \&
  Ginsberg]{Constantin2020}
{\sc \au{Constantin, Peter}, \au{Drivas, Theodore~D} \& \au{Ginsberg, Daniel}}
  \yr{2020}  \at{On quasisymmetric plasma equilibria sustained by small force}.
   \jt{arXiv preprint arXiv:2009.08860} .

\bibitem[Drevlak {\em et~al.\/}(2018)Drevlak, Beidler, Geiger, Helander \&
  Turkin]{Drevlak2018}
{\sc \au{Drevlak, M}, \au{Beidler, CD}, \au{Geiger, J}, \au{Helander, P} \&
  \au{Turkin, Y}} \yr{2018}  \at{Optimisation of stellarator equilibria with
  {ROSE}}.  \jt{Nuclear Fusion}  \bvol{59}~(1),  \pg{016010}.

\bibitem[Elbarmi {\em et~al.\/}(2020)Elbarmi, Sengupta \&
  Weitzner]{elbarmi_sengupta_weitzner_2020}
{\sc \au{Elbarmi, Elena}, \au{Sengupta, Wrick} \& \au{Weitzner, Harold}}
  \yr{2020}  \at{Charged particle dynamics near an {X-point} of a non-symmetric
  magnetic field with closed field lines}.  \jt{Journal of Plasma Physics}
  \bvol{86}~(2),  \pg{905860209}.

\bibitem[Garren \& Boozer(1991{\natexlab{{\em a\/}}})]{Garren1991a}
{\sc \au{Garren, DA} \& \au{Boozer, Allen~H}} \yr{1991{\natexlab{{\em a\/}}}}
  \at{Existence of quasihelically symmetric stellarators}.  \jt{Physics of
  Fluids B: Plasma Physics}  \bvol{3}~(10),  \pg{2822--2834}.

\bibitem[Garren \& Boozer(1991{\natexlab{{\em b\/}}})]{Garren1991b}
{\sc \au{Garren, David~Alan} \& \au{Boozer, AH}} \yr{1991{\natexlab{{\em
  b\/}}}}  \at{Magnetic field strength of toroidal plasma equilibria}.
  \jt{Physics of Fluids B: Plasma Physics}  \bvol{3}~(10),  \pg{2805--2821}.

\bibitem[Hegna(2000)]{Hegna2000}
{\sc \au{Hegna, CC}} \yr{2000}  \at{Local three-dimensional magnetostatic
  equilibria}.  \jt{Physics of Plasmas}  \bvol{7}~(10),  \pg{3921--3928}.

\bibitem[Helander(2014)]{Helander2014}
{\sc \au{Helander, Per}} \yr{2014}  \at{Theory of plasma confinement in
  non-axisymmetric magnetic fields}.  \jt{Reports on Progress in Physics}
  \bvol{77}~(8),  \pg{087001}.

\bibitem[Henneberg {\em et~al.\/}(2019)Henneberg, Drevlak \&
  Helander]{Henneberg2019}
{\sc \au{Henneberg, SA}, \au{Drevlak, M} \& \au{Helander, P}} \yr{2019}
  \at{Improving fast-particle confinement in quasi-axisymmetric stellarator
  optimization}.  \jt{Plasma Physics and Controlled Fusion}  \bvol{62}~(1),
  \pg{014023}.

\bibitem[Hirshman \& Whitson(1983)]{Hirshman1983}
{\sc \au{Hirshman, S.~P.} \& \au{Whitson, J.~C.}} \yr{1983}
  \at{Steepest‐descent moment method for three‐dimensional
  magnetohydrodynamic equilibria}.  \jt{The Physics of Fluids}  \bvol{26}~(12),
   \pg{3553--3568},  \arxiv{arXiv:
  https://aip.scitation.org/doi/pdf/10.1063/1.864116}.

\bibitem[Imbert-Gerard {\em et~al.\/}(2019)Imbert-Gerard, Paul \&
  Wright]{Imbert2019}
{\sc \au{Imbert-Gerard, Lise-Marie}, \au{Paul, Elizabeth} \& \au{Wright,
  Adelle}} \yr{2019}  \at{An introduction to symmetries in stellarators}.
  \jt{arXiv preprint arXiv:1908.05360} .

\bibitem[Jaquiery \& Sengupta(2019)]{Jaquiery2019}
{\sc \au{Jaquiery, Erin} \& \au{Sengupta, Wrick}} \yr{2019}  \at{Low-shear
  three-dimensional equilibria in a periodic cylinder}.  \jt{Journal of Plasma
  Physics}  \bvol{85}~(1),  \pg{905850115}.

\bibitem[Jorge \& Landreman(2020)]{Jorge2020b}
{\sc \au{Jorge, Rogerio} \& \au{Landreman, Matt}} \yr{2020}  \at{The use of
  near-axis magnetic fields for stellarator turbulence simulations}.
  \jt{Plasma Physics and Controlled Fusion} .

\bibitem[Jorge {\em et~al.\/}(2019)Jorge, Sengupta \& Landreman]{Jorge2019}
{\sc \au{Jorge, R}, \au{Sengupta, W} \& \au{Landreman, M}} \yr{2019}
  \at{Near-axis expansion of stellarator equilibrium at arbitrary order in the
  distance to the axis}.  \jt{arXiv preprint arXiv:1911.02659} .

\bibitem[Jorge {\em et~al.\/}(2020)Jorge, Sengupta \& Landreman]{Jorge2020}
{\sc \au{Jorge, Rogerio}, \au{Sengupta, Wrick} \& \au{Landreman, Matt}}
  \yr{2020}  \at{Construction of quasisymmetric stellarators using a direct
  coordinate approach}.  \jt{Nuclear Fusion}  \bvol{60},  \pg{076021}.

\bibitem[Landreman(2019)]{Landreman2019b}
{\sc \au{Landreman, Matt}} \yr{2019}  \at{Optimized quasisymmetric stellarators
  are consistent with the garren--boozer construction}.  \jt{Plasma Physics and
  Controlled Fusion}  \bvol{61}~(7),  \pg{075001}.

\bibitem[Landreman \& Catto(2012)]{Landreman2012}
{\sc \au{Landreman, Matt} \& \au{Catto, Peter~J}} \yr{2012}  \at{Omnigenity as
  generalized quasisymmetry}.  \jt{Physics of Plasmas}  \bvol{19}~(5),
  \pg{056103}.

\bibitem[Landreman \& Jorge(2020)]{Landreman2020b}
{\sc \au{Landreman, Matt} \& \au{Jorge, Rogerio}} \yr{2020}  \at{Magnetic well
  and mercier stability of stellarators near the magnetic axis}.  \jt{arXiv
  preprint arXiv:2006.14881} .

\bibitem[Landreman \& Sengupta(2018)]{Landreman2018a}
{\sc \au{Landreman, Matt} \& \au{Sengupta, Wrick}} \yr{2018}  \at{Direct
  construction of optimized stellarator shapes. part 1. theory in cylindrical
  coordinates}.  \jt{Journal of Plasma Physics}  \bvol{84}~(6).

\bibitem[Landreman \& Sengupta(2019)]{Landreman2019}
{\sc \au{Landreman, Matt} \& \au{Sengupta, Wrick}} \yr{2019}  \at{Constructing
  stellarators with quasisymmetry to high order}.  \jt{Journal of Plasma
  Physics}  \bvol{85}~(6).

\bibitem[Landreman {\em et~al.\/}(2018)Landreman, Sengupta \&
  Plunk]{Landreman2018b}
{\sc \au{Landreman, Matt}, \au{Sengupta, Wrick} \& \au{Plunk, Gabriel~G}}
  \yr{2018}  \at{Direct construction of optimized stellarator shapes. ii.
  numerical quasisymmetric solutions}.  \jt{arXiv preprint arXiv:1809.10246} .

\bibitem[N{\"u}hrenberg \& Zille(1988)]{Nuhrenberg1988}
{\sc \au{N{\"u}hrenberg, J} \& \au{Zille, R}} \yr{1988}  \at{Quasi-helically
  symmetric toroidal stellarators}.  \jt{Physics Letters A}  \bvol{129},
  \pg{113}.

\bibitem[Palumbo(1968)]{Palumbo1968}
{\sc \au{Palumbo, Donato}} \yr{1968}  \at{Some considerations on closed
  configurations of magnetohydrostatic equilibrium}.  \jt{Il Nuovo Cimento B
  (1965-1970)}  \bvol{53}~(2),  \pg{507--511}.

\bibitem[Plunk \& Helander(2018)]{Plunk2018}
{\sc \au{Plunk, GG} \& \au{Helander, Per}} \yr{2018}  \at{Quasi-axisymmetric
  magnetic fields: weakly non-axisymmetric case in a vacuum}.  \jt{Journal of
  Plasma Physics}  \bvol{84}~(2).

\bibitem[Plunk(2020)]{Plunk2020}
{\sc \au{Plunk, G.~G.}} \yr{2020}  \at{Perturbing an axisymmetric magnetic
  equilibrium to obtain a quasi-axisymmetric stellarator}.  \jt{Journal of
  Plasma Physics}  \bvol{86}~(4),  \pg{905860409}.

\bibitem[Rodriguez \& Bhattacharjee(2020{\natexlab{{\em
  a\/}}})]{Rodriguez2020c}
{\sc \au{Rodriguez, Eduardo} \& \au{Bhattacharjee, Amitava}}
  \yr{2020{\natexlab{{\em a\/}}}}  \at{Solving the problem of overdetermination
  of quasisymmetric equilbrium solutions by near-axis expansions: {I.
  Generalised} force balance}.  \jt{arXiv preprint arXiv:2008.04715} .

\bibitem[Rodriguez \& Bhattacharjee(2020{\natexlab{{\em
  b\/}}})]{Rodriguez2020b}
{\sc \au{Rodriguez, Eduardo} \& \au{Bhattacharjee, Amitava}}
  \yr{2020{\natexlab{{\em b\/}}}}  \at{Solving the problem of overdetermination
  of quasisymmetric equilbrium solutions by near-axis expansions. {II. Circular
  axis stellarators}}.  \jt{arXiv preprint arXiv:2008.12580} .

\bibitem[Rodriguez {\em et~al.\/}(2020)Rodriguez, Helander \&
  Bhattacharjee]{Rodriguez2020a}
{\sc \au{Rodriguez, Eduardo}, \au{Helander, Per} \& \au{Bhattacharjee,
  Amitava}} \yr{2020}  \at{Necessary and sufficient conditions for
  quasisymmetry}.  \jt{Physics of Plasmas}  \bvol{27}~(6),  \pg{062501}.

\bibitem[Sanchez {\em et~al.\/}(2000)Sanchez, Hirshman, Ware, Berry \&
  Spong]{Sanchez2000}
{\sc \au{Sanchez, R}, \au{Hirshman, SP}, \au{Ware, AS}, \au{Berry, LA} \&
  \au{Spong, DA}} \yr{2000}  \at{Ballooning stability optimization of
  low-aspect-ratio stellarators}.  \jt{Plasma Physics and Controlled Fusion}
  \bvol{42}~(6),  \pg{641}.

\bibitem[Sengupta \& Weitzner(2019)]{Sengupta2019}
{\sc \au{Sengupta, Wrick} \& \au{Weitzner, Harold}} \yr{2019}  \at{Low-shear
  three-dimensional equilibria and vacuum magnetic fields with flux surfaces}.
  \jt{Journal of Plasma Physics}  \bvol{85}~(2),  \pg{905850209}.

\bibitem[Skovoroda(2009)]{Skovoroda2009}
{\sc \au{Skovoroda, AA}} \yr{2009}  \at{Local surface equilibrium equations for
  currentless magnetic configurations}.  \jt{Plasma Physics Reports}
  \bvol{35}~(2),  \pg{99--111}.

\bibitem[Spong {\em et~al.\/}(2001)Spong, Hirshman, Berry, Lyon, Fowler,
  Strickler, Cole, Nelson, Williamson, Ware {\em et~al.\/}]{Spong2001}
{\sc \au{Spong, DA}, \au{Hirshman, Steven~P}, \au{Berry, LA}, \au{Lyon, JF},
  \au{Fowler, RH}, \au{Strickler, DJ}, \au{Cole, MJ}, \au{Nelson, BN},
  \au{Williamson, DE}, \au{Ware, AS} \& \au{others}} \yr{2001}  \at{Physics
  issues of compact drift optimized stellarators}.  \jt{Nuclear Fusion}
  \bvol{41}~(6),  \pg{711}.

\bibitem[Weitzner(2016)]{Weitzner2016}
{\sc \au{Weitzner, Harold}} \yr{2016}  \at{Expansions of non-symmetric toroidal
  magnetohydrodynamic equilibria}.  \jt{Physics of Plasmas}  \bvol{23}~(6),
  \pg{062512}.

\bibitem[Weitzner \& Sengupta(2020)]{Weitzner2020}
{\sc \au{Weitzner, Harold} \& \au{Sengupta, Wrick}} \yr{2020}  \at{Exact
  non-symmetric closed line vacuum magnetic fields in a topological torus}.
  \jt{Physics of Plasmas}  \bvol{27}~(2),  \pg{022509}.

\end{thebibliography}

\end{document}